\documentclass[useAMS,usenatbib]{mn2e}
\usepackage{ulem}
\usepackage{hyperref}
\usepackage{amsmath}
\usepackage{amssymb}
\usepackage{caption}
\usepackage{graphicx}
\usepackage{color}
\newcommand{\beq}{\begin{equation}}
\newcommand{\eeq}{\end{equation}}
\newcommand{\beqa}{\begin{eqnarray}}
\newcommand{\eeqa}{\end{eqnarray}}

\newcommand{\mnras}{$MNRAS$}
\newcommand{\apj}{$ApJ$}
\newcommand{\apjl}{$ApJ$}

\newcommand{\aap}{$A\&A$}

\newcommand{\apjs}{$ApJS$}

\title[Gyro-cooling of electrons in magnetic loops]{On the spatial distribution of electron energy loss due to gyro-cooling in hot-star magnetospheres}
\author[B. Das \& S. Owocki]{B.\ Das$^{1,2}$\thanks{E-mail:
Barnali.Das@csiro.au} and 
S.\ P.\ Owocki$^{1}$
\\
$^1$Department of Physics and Astronomy, University of Delaware, 217 Sharp Lab, Newark, Delaware, 19716, USA\\
$^2$CSIRO Space and Astronomy, PO Box 1130, Bentley WA 6102, Australia\\
}
\begin{document}
\include{aas_macros}

\date{}

\pagerange{\pageref{firstpage}--\pageref{lastpage}} \pubyear{2002}

\maketitle

\label{firstpage}

\begin{abstract}
Hot magnetic stars often exhibit incoherent circularly polarized radio emission thought to arise from gyro-synchrotron emission by energetic electrons trapped in the circumstellar magnetosphere. 
Theoretical scalings for electron acceleration by magnetic reconnection driven by centrifugal breakout match well the empirical scalings for observed radio luminosity with both the magnetic field strength and the stellar rotation rate. 
This paper now examines how energetic electrons introduced near the top of closed magnetic loops are subsequently cooled by the energy loss associated with their gyro-synchrotron radio emission. 
For sample assumed distributions for energetic electron deposition about the loop apex, we derive the spatial distribution of the
radiated energy from such``gyro-cooling''. For sub-relativistic electrons, we show explicitly that this is independent of the input energy, but also find that even extensions to the relativistic regime still yield a quite similar spatial distribution. % of emission.
However, cooling by coulomb collisions with even a modest ambient density of thermal electrons can effectively quench the emission from sub-relativistic electrons, indicating that the observed radio emission likely stems from relativistic electrons that are less affected by such collisional cooling.
The overall results form an initial basis for computing radio emission spectra in future models that account for such cooling and multimode excitation about the fundamental gyro-frequency.
Though motivated in the context of hot-stars, the basic results here could also be applied to gyro-emission in any  dipole magnetospheres, including those of ultra-cool dwarfs and even (exo)-planets.
\end{abstract}


\section{Introduction}
Hot magnetic stars with moderately rapid rotation show incoherent, circularly polarized radio emission, thought to arise from gyro-synchrotron emission of energetic electrons trapped in magnetic loops
\citep[e.g.][]{1987ApJ...322..902D,andre1988,2004A&A...418..593T}.
Empirical analyses show that the radio emission depends on  both the magnetic field strength and the stellar rotation rate
\citep{leto2021,shultz2021},
with a scaling that is well explained by a model in which the electrons are energized by magnetic reconnection
events that arise from centrifugal breakout (CBO) of plasma trapped in the rotating magnetosphere \citep{owocki2022}.
%(Owocki et al. 2022).
Building thus on a scenario in which energetic electrons are introduced around the apex of closed magnetic loops, the present paper examines
how the energy lost to gyro-synchrotron radio emission cools the electrons, and how this affects the spatial and spectral distribution of the observed
radio emission.

%Hot luminous, massive stars of spectral type O and B have dense, high-speed, radiatively driven stellar winds
%\citep{cak1975}.
In the subset ($\sim$10\%; \cite{2017MNRAS.465.2432G,2019MNRAS.483.2300S}) of OBA stars with strong ($>\,100$\,G; \cite{2007AA...475.1053A,2019MNRAS.482.3950S}), globally ordered (often significantly dipolar; \citet{2019A&A...621A..47K}) magnetic fields, the trapping of stellar wind outflow by closed magnetic loops leads to the formation of a circumstellar {\it  magnetosphere} \citep{petit2013}.
Because of the angular momentum loss associated with 
their relatively strong, magnetised wind 
\citep{ud2009}, magnetic O-type stars are typically
slow rotators,  with trapped wind material falling back on a dynamical timescale, giving what's known as a ``dynamical magnetosphere" (DM).

But in cooler magnetic stars (spectral type B or even A), the relatively weak stellar winds imply longer spin-down times, and so a significant fraction that still retain a moderately rapid rotation.
For cases in which the associated Keplerian co-rotation radius $R_{\rm K}$ lies within the Alfv\'{e}n radius $R_{\rm A}$ that characterises the maximum height of closed loops, the rotational 
support leads to formation of a ``{\it centrifugal magnetosphere}'' (CM).
Recent work has shown how the {\it centrifugal breakout} (CBO) from such CM's plays a key role in both their Balmer line emission
\citep{shultz2020,owocki2020},
%\citep{2020MNRAS.499.5379S,2020MNRAS.499.5366O}, 
as well as their incoherent radio emission \citep{owocki2022}.

As a basis toward developing predictive models of such radio emission, the analysis here assumes that repeated CBO-driven magnetic reconnection events seed a quasi-steady, gyrotropic population of energetic electrons around the tops of underlying closed magnetic loops.
As in the standard scenario \citep[e.g.][]{leto2021}, the spiraling of these energetic electrons as they mirror between the opposite footpoints of the loop leads to the gyro-synchrotron emission of the observed radio.

In the past, \citet{2004A&A...418..593T} presented a three-dimensional numerical model to calculate gyro-synchrotron radio emission from hot magnetic stars using a set of free parameters related to the stellar magnetospheres. This model is based on the scenario proposed by \citet{andre1988},
wherein electrons are accelerated to relativistic energies at  a current sheet that lies at the magnetic equator near the Alfv\'en radius; these electrons travel towards the star following the magnetic field lines and emit non-thermal radio emission.
It is assumed that the non-thermal electrons have a power-law distribution in energy, but maintains an isotropic distribution at all points of the `middle magnetosphere' \citep[e.g.][]{2004A&A...418..593T}. This model has been used in several subsequent works \citep[e.g.][etc.]{2006A&A...458..831L,leto2012,leto2021} to estimate different stellar parameters via comparison with observed radio emission. 

In the present study, we investigate how the energy emitted by the non-thermal electrons depends on the details of the energy `deposition' in the magnetosphere. 
%several new aspects of the phenomenon of magnetospheric radio emission by hot magnetic stars, such as the dependence of the spatial distribution of the emitted energies on the locations of the sites of production of energetic electrons, and also, the role of pitch angle on determining the radio spectra. 
While doing this, we self-consistently consider the time-evolution of the electron pitch-angles due to loss of energy via radiation (i.e. the effect of non-conservation of magnetic moment). Our work is motivated from 
recent empirical results that the current sheet near the Alfv\'en radius  cannot be the main source of non-thermal electron production in stars with centrifugal magnetospheres
\citep{leto2021,shultz2021}; instead
CBO events are more favoured candidates to explain the observed relation of radio luminosity with stellar rotation and other parameters
\citep{shultz2021,owocki2022}.
Our ultimate goal is to understand the physical process(es) that produce(s) the non-thermal electrons in hot-star magnetospheres. This work is the first step towards that goal of connecting the observed properties of radio emission with that of the phenomenon responsible for generating the relativistic electrons.

%The central new feature of our analysis is now to account self-consistently for the energy loss associated with this radio emission, i.e. the back-effect of ``gyro-cooling'' on such radio emission.

As detailed in the next section (\S 2.1), for magnetic field strengths $\gtrsim 100$\,G that are associated with B-star CM's, the electron cooling time due to radiation of energy is less than a day,  and so comparable to or shorter than the characteristic timescales for wind outflow and filling the CM\footnote{
Within the general lack of strong variability in the observed incoherent radio flux \citep[e.g.][etc.]{2000AA...362..281T, das2021}, this supports the evidence from similarly steady H$\alpha$ and continuum flux observations \citep{shultz2020,owocki2020} that CBO occurs not in sporadic, large-scale events, but rather through a quasi-steady stream of smaller outbursts.}.
After reviewing the general features of propagation and mirroring of electrons along a closed loop (\S 2.2), we derive (\S 3.1) the basic coupled differential equations
%ODE's 
for loss rate of electron energy, and the associated evolution of its magnetic moment. 
For the common case that the electron propagation time is much smaller than its cooling time, solutions show (\S 3.2) that the spatial distribution of radio emission along the loop follows a distinctive form (Figure \ref{fig:etotvmu-3k}), with a narrow peak near the loop apex, and broader wings extending down to sharp cutoffs due to truncation from the underlying star.
We then explore models with energy deposition that have gaussian distributions in radius and latitude (\S 4.1),
deriving the associated spatial distribution of energy lost due to gyrocooling
%emitted intensity 
(\S 4.2).
%\sout{For the simplest excitation model in which radio emission occurs predominantly at the fundamental gyro-frequency, we derive (\S 4.3) sample synthetic spectra for various assumptions for distribution of CBO deposition of energetic electrons.}
Following a brief analysis (\S 4.3) of the cooling effect of coulomb collisions with thermal electrons, 
we conclude (\S 5) with a brief summary and outline for future work.
The appendix presents generalized equation forms for relativistic electrons.

\section{Background}

\subsection{Gyro-synchrotron cooling}

For a potentially relativistic electron with pitch-angle $\alpha$ in gyration about a local magnetic field, the associated synchrotron power emitted is given by \citet[][their eq. (5.37)]{ConRan16},
\beq
 P  = 2 \sigma_T \beta^2 \gamma^2 c U_B 
 \, \sin^2 \alpha
\, ,
\label{eq:Pdef}
\eeq
where
$U_B \equiv B^2/8 \pi$ is the (cgs) energy density of the magnetic field $B$;
$\sigma_T = 0.67 \times 10^{-24}$ cm$^2$ is the Thomson cross section for electron scattering;
$\beta \equiv v/c$ is the ratio of electron speed $v$ to the speed of light $c$; and
$\gamma \equiv 1/\sqrt{1-\beta^2}$ is the associated relativistic energy factor.

For electron mass $m_e$, the associated electron kinetic energy is KE$= (\gamma -1 )m_e c^2$.
Averaging the power loss over a given pitch-angle distribution $\left  < P \right > \sim \left <\sin^2 \alpha \right >$,
we can then define an electron energy-loss time  as $t_e \equiv$ KE/$\left <P \right >$, which,
after some manipulation, can be shown to scale as
\beqa
~~~~~~~~~~~~~~~
%t_e &=& \frac{16 \pi m_e c}{\sigma_T} \,  \frac{\sin^2 \alpha}{1+\gamma} \, B^{-2} 
t_e &=& \frac{4 \pi m_e c}{\sigma_T \,  (1+\gamma) B^2 \left < \sin^2 \alpha \right >} 
\nonumber
\\
&=& \frac{5.16 \times 10^8  {\rm s}}{(1+\gamma) \left < \sin^2 \alpha \right >} \, 
      \left ( \frac{1 \, {\rm G}}{B} \right )^2
\, .
\label{eq:teloss}
\eeqa

\noindent
For non-relativistic electrons (i.e. with $\gamma \gtrsim 1$) with a given pitch angle distribution, 
note that this cooling time depends {\it only} on the magnetic field strength, and not, e.g., on the electron speed or energy.

For the simple case of a gyrotropic distribution (for which 
$\left < \sin^2 \alpha \right > = 2/3$) and a canonical field $B=1$G,  the latter relation in (\ref{eq:teloss}) shows that the loss time 
%{\color{blue} for non-relativistic electrons ($\gamma \appro 1$) }
is more than a decade (recalling that 1\,yr\,$\approx 3 \times 10^7$\,s).
%\sout{That of course is much longer than any evolution time of flares, coronal mass ejections, or associated radio bursts, indicating that gyro-synchrotron emission is not a significant cooling mechanism for such solar processes.}

However, in the context of massive-star magnetospheres, this loss timescale can be much shorter.
Analyses by \citet{shultz2021} and \citet{owocki2020} have shown that H-alpha emission requires a field at the Kepler radius $ B_K \approx 100$G,
implying a cooling time for non-relativistic electrons, $t_e \approx 5.1 \times 10^4 {\rm s}  \approx 0.6$\,d.
The implication is then that the observed quasi-steady circularly polarized gyro-synchrotron radio emission must be replenished by many small centrifugal breakouts that occur on such a timescale of order a day or less.

On the other hand, this loss time is generally much {\it longer} than the typical propagation time of such energetic electrons across loops.
As an example, for 25 keV electrons with speed $v \approx c/3$,
the characteristic advection time across of loop of apex radius  $r_a$ is
\beq
t_a \equiv \frac{r_a}{v} 
\approx 100 \, {\rm s} \, \frac{r_a}{10^{12} \, {\rm cm}} 
\, .
\label{eq:tadef}
\eeq
Comparison with eqn.\ (\ref{eq:teloss}) shows that for a loop with $r_a \approx 10^{12}$cm and $B_a \approx 100$G, $t_a \ll t_e$.
As detailed below, this implies that electrons will generally mirror many times across a dipole loop as they gradually lose energy. \\\
%electrons will mirror about a  hundred times before they cool.  

%These general scalings and estimates provide the groundwork for the further, more detailed analyses below.

%{\color{blue}
%Blue: Stan suggested briefer review on mirroring, to include in section 2 instead of current section 3.

\subsection{Magnetic mirroring}

An electron with energy $E$ and pitch angle $\alpha$ gyrating around field of local strength
$B$ has a magnetic moment given by
\beq
p_m\equiv E \sin^2 \alpha/B
\, .
\label{eq:pmdef}
\eeq
Under the common assumption of fixed energy $E$,
this magnetic moment is also conserved.
For electrons with pitch angle $\alpha_a$ at the apex radius $r_a$ of a loop with apex field strength $B_a$, the pitch angle will thus become perpendicular ($\sin \alpha = 1$) at a mirror radius $r_m$, with field strength  
\beq
B_m = \frac{B_a}{\sin^2 \alpha_a}
\, .
\label{eq:Bm}
\eeq
For a simple dipole field with spatial scaling $B \sim \sqrt{1+3\mu^2}/r^3$ in radius $r$ and colatitude cosine  $\mu = \sqrt{1-r/r_a}$, 
this leads to a sixth-order polynomial for $r_m$,
\beq
\left ( \frac{r_m}{r_a} \right )^6 = 
\sin^4 \alpha_a \, (4 - 3 r_m/r_a)
\, .
\label{eq:rmra6}
\eeq
%\sout {The top panel of Figure \ref{fig:rmir} plots the full solution for the ratio $r_m/r_a$ vs. $\alpha_a$ (solid curve), comparing this with the simple approximate solution $r_m/r_a \approx \sin^{2/3} \alpha_a$ (dashed curve). The bottom panel shows the corresponding co-latitude cosine $\mu_m = \sqrt{1-r_m/r_a}$.}
The simple approximation
  $r_m/r_a \approx \sin^{2/3} \alpha_a$ is
  within a few percent of the full solution
  except for the small pitch angle regime. For the analysis presented in the subsequent sections, we have used the full solution.

 \section{Cooling along loop}
 
 \subsection{Dimensionless ODE for energy loss}\label{subsec:gov_eqns}

 More generally, the time variation of energy $dE/dt$ associated with the gyro-synchrotron radiation leads to an associated change in the magnetic moment,
 %the following equation:
 \begin{align}
     \frac{dp_m}{dt}&=\frac{1}{B}\frac{dE}{dt}
     \, .
\label{eq:dpmdt}
 \end{align}
 For an electron with speed $v$, energy $E=m_e v^2/2$, and pitch angle $\alpha$,  the gyro-emission power $P$ from eqn. (\ref{eq:Pdef}) gives for the time change of energy,
 \beq
 \frac{dE}{dt} 
% = v \cos \alpha \frac{dE}{ds}
  =  -  P 
%   \approx - \frac{\sigma_T}{3 \pi c}   v^2 B^2   \sin^2 \alpha
  \approx - \frac{ \sigma_T}{2\pi m_e c}  E B^2   \sin^2 \alpha
  =  - \frac{ \sigma_T}{2 \pi m_e c}  p_m B^3    , 
\label{eq:dEdt}
\eeq
where the second equality applies for the non-relativistic case $\gamma \approx 1$, and the final equality
uses the definition of magnetic moment $p_m = E \sin^2 \alpha/B$  to eliminate $\alpha$ in favor of $B/E$.

The initial analysis here focuses on this sub-relativistic case because then the relative cooling becomes independent of the electron energy, depending only on the magnetic field strength $B$ and electron pitch angle $\alpha$.
Appendix A derives the generalized equations for relativistic electrons.

Defining apex-scaled variables for energy $e \equiv E/E_a$ and magnetic field $b \equiv B/B_a$, along with scaled  magnetic moment 
$p \equiv p_m B_a/E_a$, eqns. (\ref{eq:dEdt}) and (\ref{eq:dpmdt}) can be recast as a coupled system of two first-order, dimensionless ODE's,
\beq
\boxed{ \frac{de}{dt} =  - k \, p  \, b^3}
\, ,
\label{eq:dedt}
\eeq
and
\beq
%\boxed{ \frac{dp}{dt} =  - k \,  p \, b^2}
%{\color{blue}%Barnali RED
\boxed{ \frac{dp}{dt} = \frac{1}{b}\frac{de}{dt}}
%}
\, ,
\label{eq:dpdt}
\eeq
where the dimensionless constant $k$ is given by
\beq
k \equiv  
\left [
\frac{\sigma_T  B_a^2}{ 2 \pi m_e c} 
 \right ]
 \, \frac{ r_a}{v_a} 
 \equiv \frac{t_a}{t_c}
\, .
\label{eq:kdef}
\eeq
In the latter equality, $t_a  \equiv r_a/v_a$ is a characteristic propagation time,  used now to scale the time variable, $t \rightarrow t/t_a$.
The cooling constant $k$ is thereby cast as the ratio of this to a characteristic cooling time $t_c$, given 
by the inverse of the square bracket term
\footnote{This is closely related to the electron energy loss time for a specific pitch angle, as given by eq.\ (\ref{eq:teloss}).}.

For typical values $r_a = 10^{12}$\,cm, $B_a = 100$\,G, $E_a=25$\,keV,  and $v_a = c/3$,  
we find this cooling constant is quite small, $k \approx 0.002$.
For initial pitch angles that  are not too small,  with thus mirroring not too far below the loop apex,
only a small fraction of the particles energy is lost per mirror cycle.
But the cubic scaling of cooling with  magnetic field, and its steep increase inward, means that
lower pitch angles can lose a significant fraction of their energy near their mirror point.

The apex-scaled dipole field  $b$ can be written as a function of the co-latitude cosine $\mu$,
\beq
b(\mu) \equiv \frac{B (\mu)}{B_a} = \frac{\sqrt{1+3 \mu^2}}{(1-\mu^2)^3}
\, .
\label{eq:bdef}
\eeq
A key to proceeding thus regards the time variation of this co-latitude variable, $d\mu/dt$. 
For electron speed $v$, movement along the field line coordinate $s$ is given by
\beq
v \cos \alpha = \frac{ds}{dt} 
= \frac{ r (d\theta/dt)}{b_\theta} 
= r_a \sqrt{1+3 \mu^2} \, \frac{d\mu}{dt}  \, ,
\eeq
where $b_\theta = \sqrt{1-\mu^2}/\sqrt{1+3\mu^2}$ is latitudinal projection of the unit vector along the dipole 
field, for which also $r = r_a (1-\mu^2)$.
 Using eqn.\ (\ref{eq:kdef}) and the above definitions of scaled energy $e$ and scaled magnetic moment $p$, 
 we can write a third dimensionless time equation that must be solved,
\beq
\boxed{\frac{d\mu}{dt} 
%= \frac {v t_c \cos \alpha}{r_a \sqrt{1+3 \mu^2}} 
= -{\rm Sign}(\cos \alpha) \, \frac {\sqrt{e-p \, b}}{ \sqrt{1+3 \mu^2}} 
}
\, ,
\label{eq:dmudt}
\eeq
where the sign function allows one to keep track the directionality of propagation as the electron mirrors across the loop.

In this approach, one solves the coupled system of 3 ODE's (\ref{eq:dmudt}), (\ref{eq:dedt}), and (\ref{eq:dpdt}) for the
associated dependent variables $\mu$, $e$, and $p$ as a function of the independent variable, the scaled time $t$.
Appendix A gives the correponding relativistic forms, 
(\ref{eq:relmu}, 
(\ref{eq:energy_eqn} 
and
(\ref{eq:p_eqn}), wherein the curly bracket factors represent the corrections from the sub-relativistic expressions here.

{
\begin{figure}
    \includegraphics[width=0.49\textwidth]{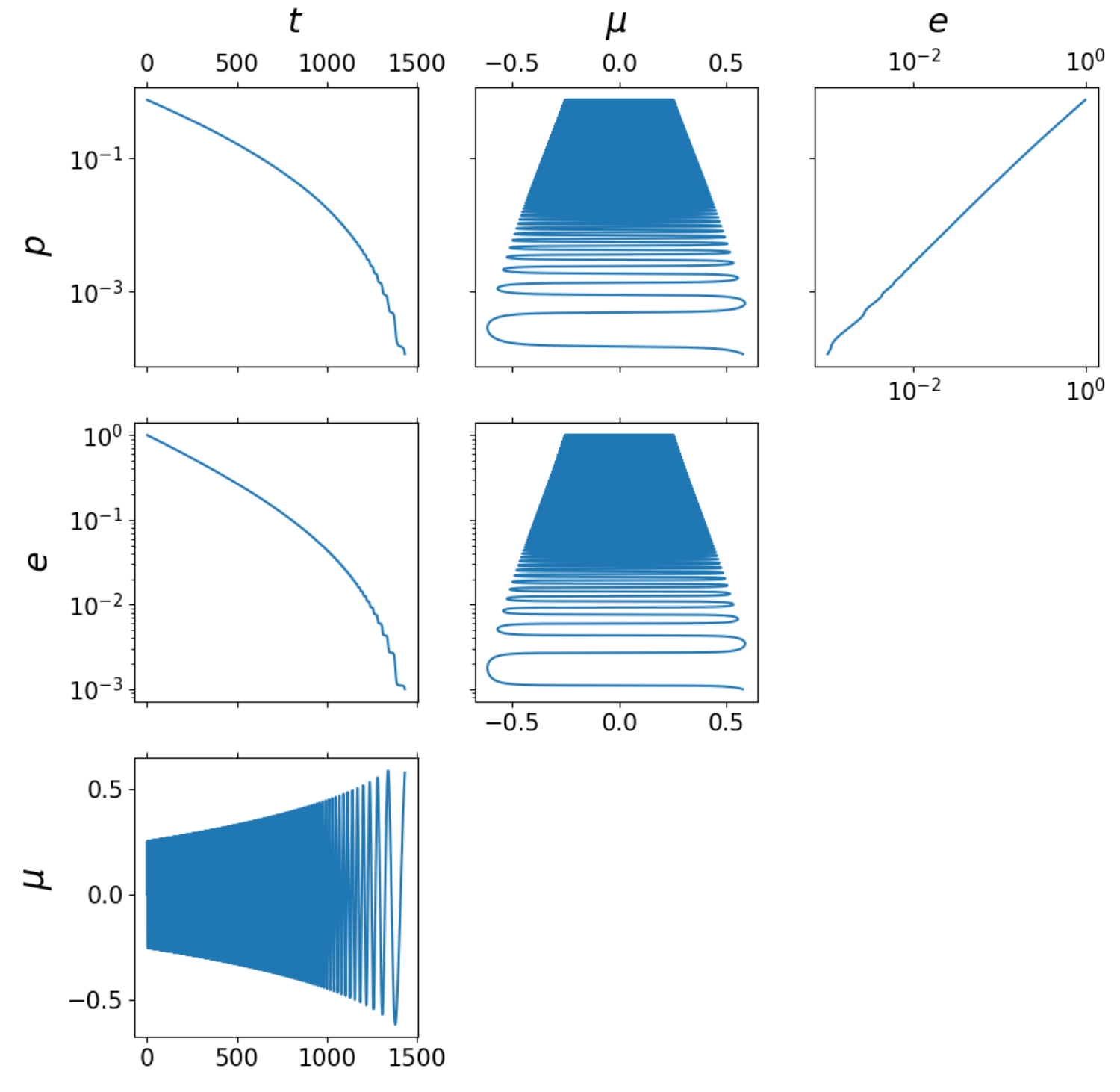}
    \caption{Solutions to the set of equations described in \S\ref{subsec:gov_eqns}. The independent variable is the dimensionless time $t$. The particle starts at the loop apex ($\mu=0$) with an initial pitch angle of $60^\circ$. For the definitions of $t$, $\mu$, $e$ and $p$, refer to \S\ref{subsec:gov_eqns}.\label{fig:sols}}
\end{figure}

\begin{figure}    %\includegraphics[width=0.45\textwidth]{delta_e_vs_mu_alpha_60deg_k_0.002_no_sum.png}
\includegraphics[width=0.45\textwidth]{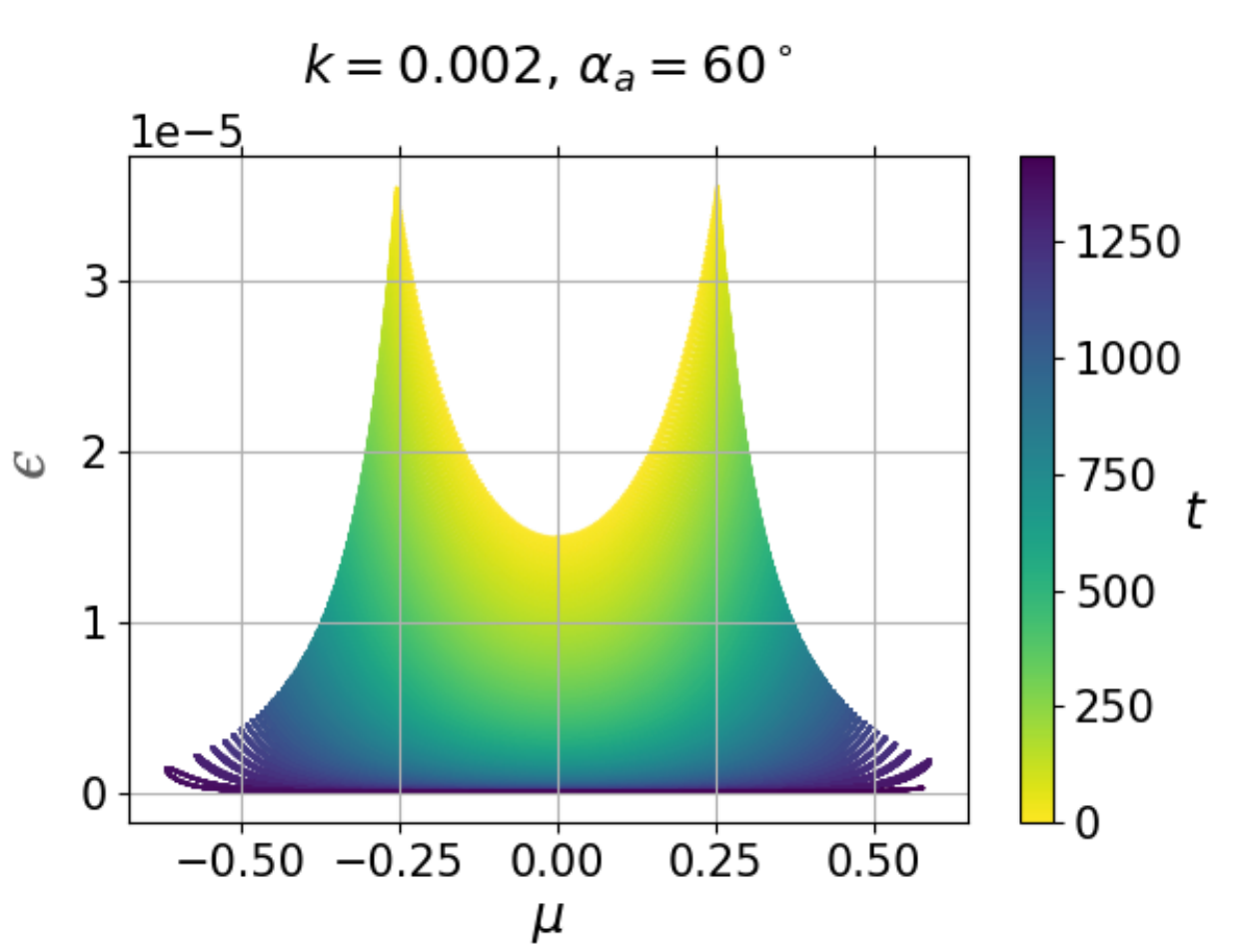}
    \caption{Color-coded time variation of energy emission $\epsilon$ plotted vs.\ co-latitude $\mu$, for the standard case with apex pitch angle $\alpha_a=60^o$ and cooling parameter $k=0.002$.}
\label{fig:devmu-nosum}
\end{figure}

\begin{figure}
\includegraphics[width=0.45\textwidth]{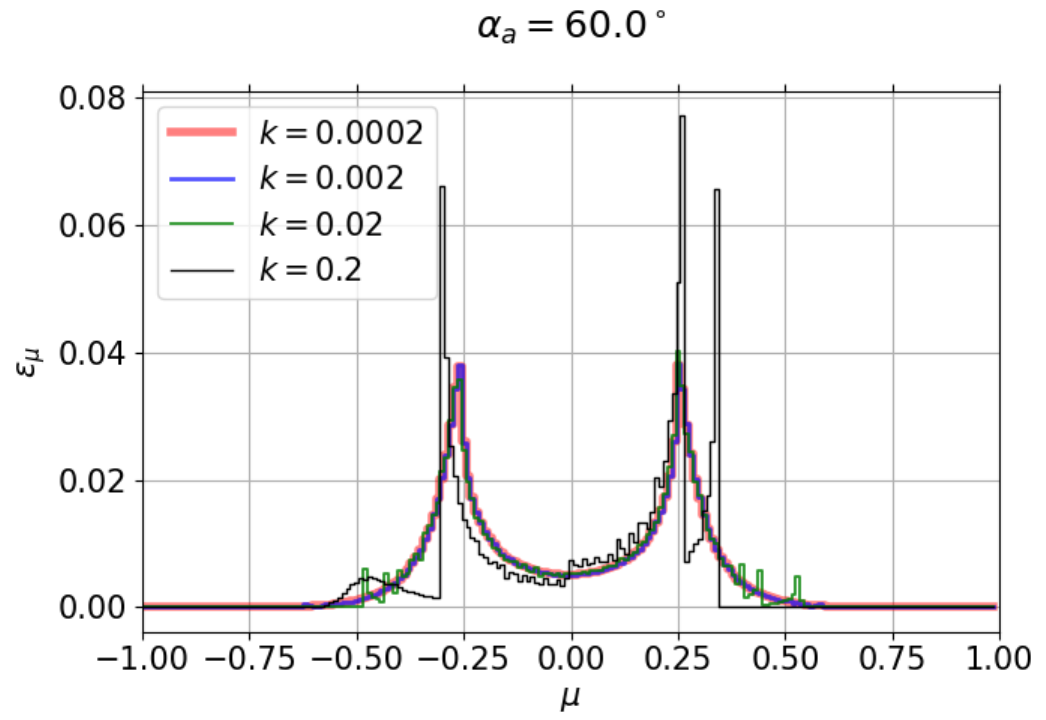}
    \caption{For the sample initial pitch angle case $\alpha_a=60^o$, the distribution of total energy emission $\epsilon_\mu$ in co-latitude $\mu$, for the four labeled values of the cooling parameter $k$.}
\label{fig:devmu-4k}
\end{figure}

\begin{figure}    %\includegraphics[width=0.45\textwidth]{delta_e_vs_mu_k_0.002_varying_alpha_normalized.png}
\includegraphics[width=0.45\textwidth]{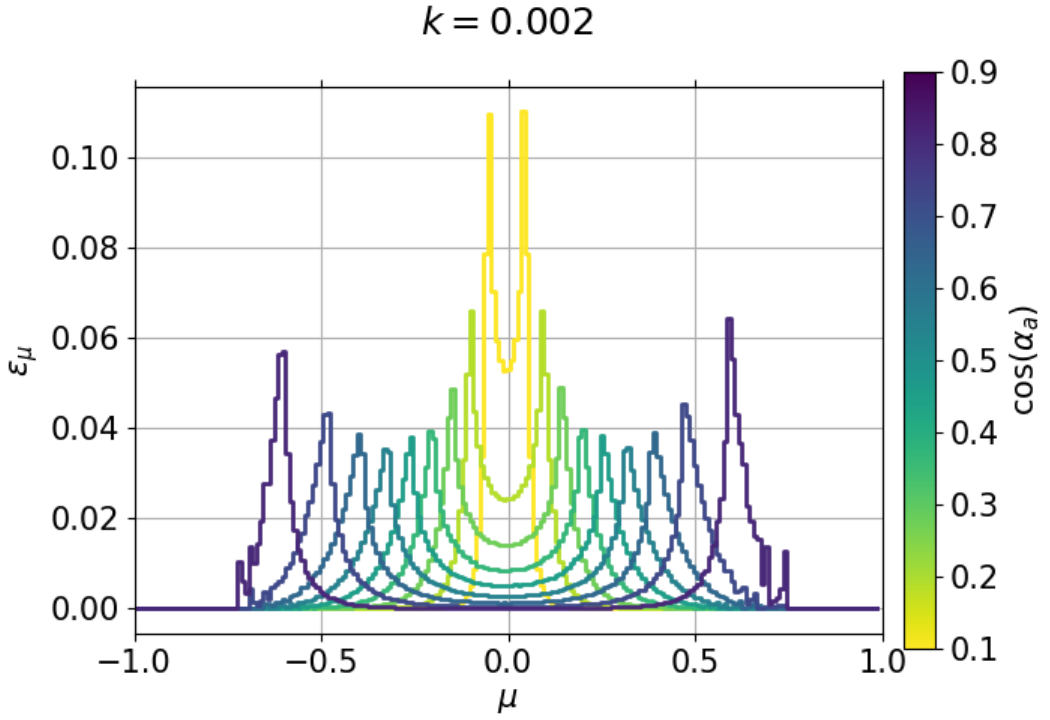}
    \caption{Total energy emission $\epsilon_\mu$ versus co-latitude $\mu$ for the labeled range of $\cos \alpha_a$.}
\label{fig:devmu-9alpha}
\end{figure}

\begin{figure}
\includegraphics[width=0.45\textwidth]{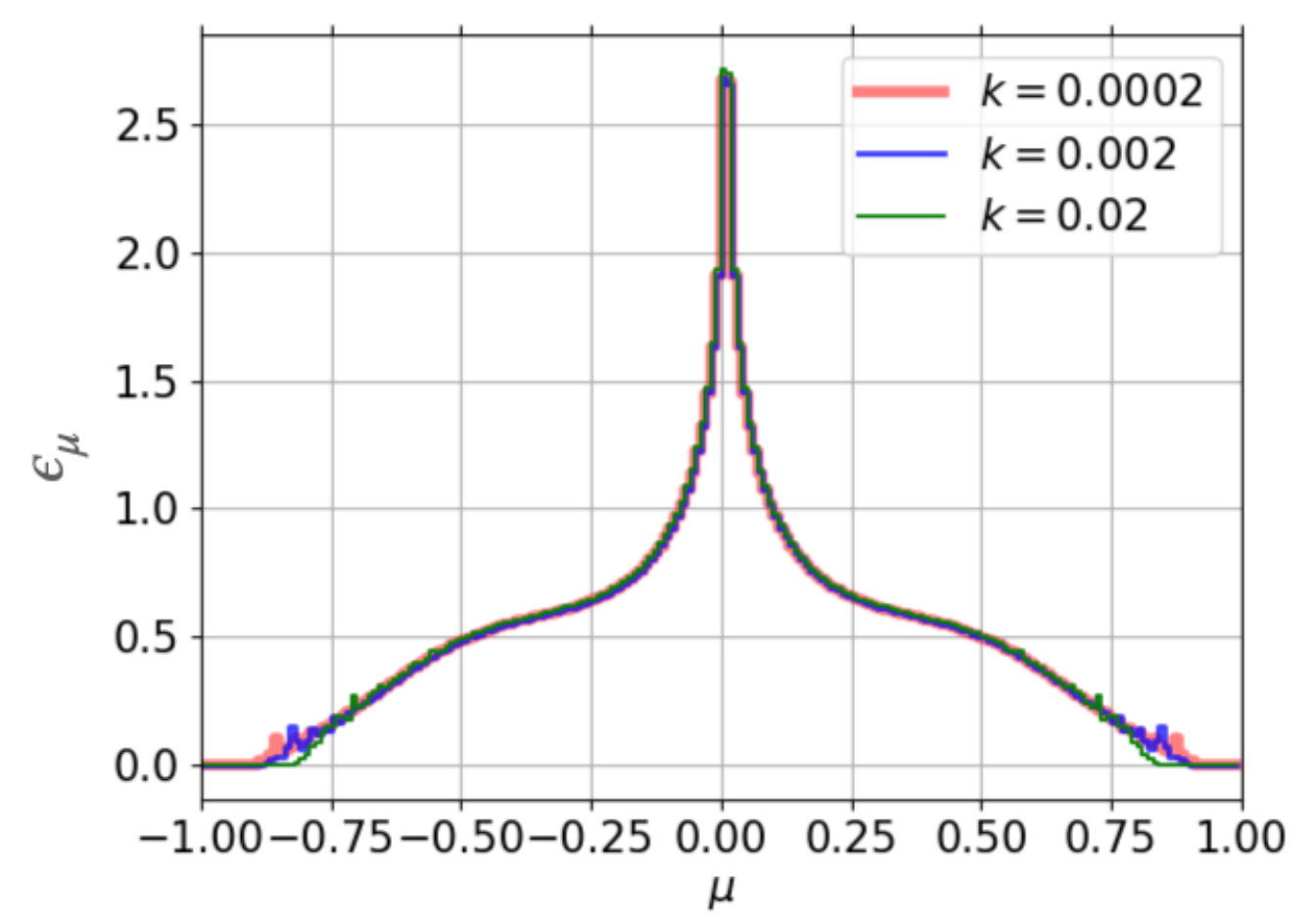}
    \caption{The cumulative energy emission for an initially gyrotropic pitch angle distribution, overplotting $\epsilon_\mu$ versus co-latitude $\mu$ for three small values of the cooling constant. The curves are normalized such that the total energy lost, as reflected by the area under the curves, is unity.}
\label{fig:etotvmu-3k}
\end{figure}

{
%\color{blue}
\subsection{Energy deposition at loop apex}}
\label{subsec:sols}
In our basic scenario here, we assume electrons are introduced at a dipole loop apex with fixed energy and pitch angle $\alpha_a$, giving then
initial conditions $\mu(0) = 0$, $e(0)=1$, and $p(0) = \sin^2 \alpha_a$. 
%\sout{Let us consider the case where $\alpha_a = 60^\circ$, and the cooling constant $k=0.002$. We }
From this inital condition, we use the \texttt{odeint} function from \texttt{scipy.integrate} \citep{2020SciPy-NMeth}} to evolve the system of ODE's (\ref{eq:dedt}), (\ref{eq:dpdt}), and (\ref{eq:dmudt})
in time until the energy retained by the electron drops below
0.1\% of its initial energy.

%\sout{At this stage, we do not yet consider the star (thus the case that the electron may be lost due to hitting the stellar surface is not considered).

%\sout{Figure \ref{fig:sols} shows the solutions obtained for this case. The bottom \sout{most} panel clearly shows the consequence of non-conservation of magnetic moment, which is that the co-latitude at which the electron mirrors, changes with time. This implies that due to the gyro-cooling effect, an electron will be able to penetrate deeper and deeper into the stellar magnetosphere as it mirrors back and forth along the magnetic field line.
%}
%}

{
%\color{blue} 
Figure \ref{fig:sols} shows solutions for a representative case with $k=0.002$ and $\alpha_a =60^o$.
The upper two panels of the left column plot the temporal decrease in the magnetic moment and energy, while the bottom panel shows the
temporal variation of the co-latitude (due to magnetic mirroring).
%mirroring of the co-latitude. 
Note how the non-conservation of magnetic moment leads to a systematic increase in the mirror co-latitude, from its initial $|\mu_{min}| \approx 0.24$ to a maximum $\mu_{max} \approx 0.55$, corresponding to a minimum radius $r_{min}/r_a \approx  1-\mu_{max}^2 \approx 0.7 $.
For any loop with apex $r_a/R_\ast > 1/0.7 \approx 1.4$, electrons with this initial pitch angle $\alpha_a = 60^o$ will thus have their full energy dissipated without interaction with the underlying star.
We thus defer consideration of such stellar interaction dissipation to the fuller models
below.

The central column of Figure \ref{fig:sols} shows the associated evolution of magnetic moment and energy with co-latitude. The latter in particular forms the basis for deriving the spatial distribution of energy loss along the loop.
Figure \ref{fig:devmu-nosum} plots the energy emission $\epsilon$ versus co-latitude $\mu$, color coded by the deposition time.

By binning the energy loss over a grid in co-latitude, Figure \ref{fig:devmu-4k} plots the total {\it distribution} of emission $\epsilon_\mu$ in co-latitude $\mu$, comparing now results for the various labeled values of the cooling parameter $k$.
Note that results for the smallest $k$ are all very similar, since they all represent the cumulative effects of energy loss over many mirror cycles;
in contrast, for the case with cooling constant $k=0.2$ approaching unity, the relatively stronger energy loss within any given mirror cycle gives a much more irregular distribution.

For the standard cooling parameter $k=0.002$, Figure \ref{fig:devmu-9alpha} overplots the energy emission $\epsilon_\mu$ versus co-latitude $\mu$ for a range of initial pitch angles ranging from $\cos (\alpha_a) =0.1$ to 0.9.
The peak energy emission again occurs near the mirror point for each initial pitch angle, with a narrow spread.
But these mirror points range from near the loop apex for oblique pitch angles (e.g., $\cos \alpha_a = 0.1$) to much further down the loop for more field-aligned cases (e.g., $\cos \alpha_a = 0.9$).

Figure \ref{fig:etotvmu-3k} plots the total energy emission for a {\it gyrotropic} distribution of initial pitch angles, assuming  the labeled small values for the cooling parameter $k$.
The close overlap for all 3 cases shows that this spatial distribution of energy is not sensitive to the exact cooling parameter, as long as it is significantly below unity, and so allows for gradual cumulative energy loss over many mirror cycles.

The strong peak around $\mu=0$ reflects the strong energy emission near the loop apex from electrons with nearly orthogonal initial pitch angles $\alpha_a \approx 90^o$, which, due to the relatively large solid angle, are relatively more numerous. But more field-aligned pitch-angles contribute to significant energy emission further down the loop, here extending to a maximum colatitude
$|\mu_{max}|$ that depends on just how small the cooling parameter $k$ is.

\vspace{0.1in}

\section{Models with extended spatial deposition of energetic electrons}

Building upon this analysis of the idealized case that energetic electrons are introduced at the apex of a single dipole loop, let us next generalize this to more realistic models in which the electron source is distributed over a range of loop heights and positions around the loop apex.

\subsection{Gaussian deposition in co-latitude}

We first consider the case in which the initial energy deposition along a given loop follows a gaussian form that is still centered on the loop apex $\mu=0$, but with co-latitude dispersion $\sigma_\mu$,
\beq
e_\mu = 
C_\mu
\exp\left(-\frac{\mu^2}{2\sigma_\mu^2}\right)
%\frac{\exp[{-(\mu/\sigma_\mu)^2}/2]}{\sqrt{2  \pi}\sigma_\mu \, \rm erf (\mu_\ast/\sqrt{2}\sigma_\mu)}
\, ,
\label{eq:dedmu_sigmu}
\eeq
where for a loop with apex radius $r_a$, $C_\mu$ is a normalization constant that ensures that the total energy ($\int e_\mu d\mu$) is unity ($e=1$) when integrated from the loop apex ($\mu=0$) to the stellar base ($\mu = \pm\mu_\ast \equiv \pm\sqrt{1-(R_\ast/r_a)}$).

\begin{figure}
\includegraphics[width=0.45\textwidth]{
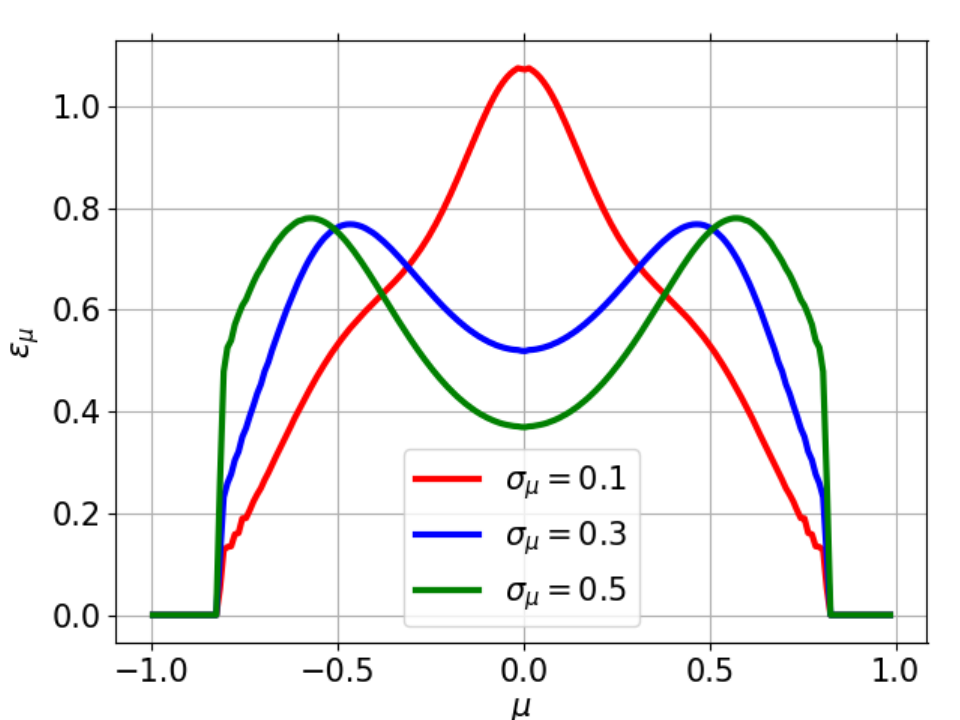}
\includegraphics[width=0.45\textwidth]{
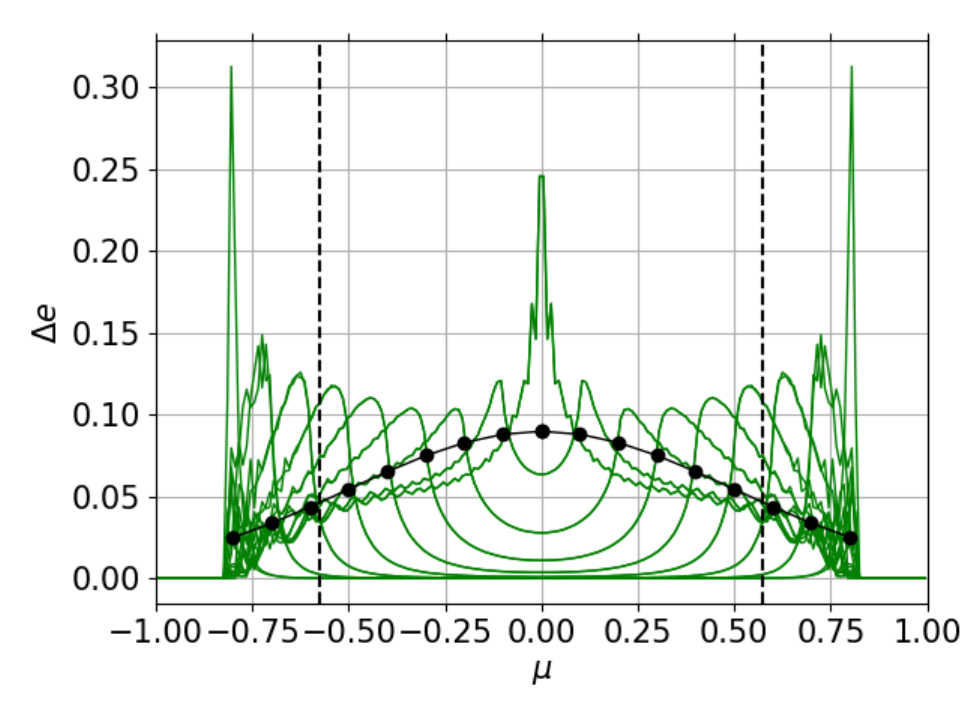}
    \caption{Top: For a sample loop with apex at $r_a = 3 R_\ast$, the energy loss $\epsilon_\mu$ versus co-latitude $\mu$ for latitudinal dispersions $\sigma_\mu = 0.1$, 0.3, and 0.5.
    In all cases, the total energy deposited $e=1$, but due to losses to the underlying star (as indicated by the sharp cutoffs at large $|\mu|$), the total emission is slightly less than unity, viz. $\left <\Delta e \right > = 0.986$, 0.985, and 0.956 for $\sigma_\mu = 0.1$, 0.3, and 0.5, respectively.
    Bottom:
    %To explain the double-peak away from loop apex 
    For the large dispersion case $\sigma_\mu=0.5$,
    the green curves break down the emission from a subset of energy distributions with peaks show by the bold black dots.
%    The dark green curve plots  
The sum of this subset forms the double peaks at location marked  here by the vertical dashed lines.
The area under each curve represents the total energy emitted corresponding to the energy deposition at a given $\mu$. 
}
\label{fig:dedmu_sigmu}
\end{figure}

For a loop with $r_a=3 R_\ast$, the top panel of Figure \ref{fig:dedmu_sigmu} compares the associated cumulative energy emission $\epsilon_\mu$ vs. co-latitude $\mu$ for cases with $\sigma_\mu = 0.1$, 0.3, and 0.5.
For the small dispersion case $\sigma_\mu = 0.1$, the cumulative energy loss still has its peak at the loop apex, though with a broader range than found for models with energy deposition confined to just the loop apex.

For intermediate dispersion case $\sigma_\mu=0.3$, the central emission is flatter, with even a modest dip around $\mu=0$.

Indeed, for the largest dispersion case $\sigma_\mu = 0.5$, the central emission shows an even deeper local minimum, even though that is still where the deposition is greatest.
Instead there are two distinct peaks located at $|\mu| \approx 0.5$, distinctly below the loop apex, where the energy deposition is highest.
%at $r \approx 0.75 r_a$
%co-latitudes closer toward the poles,

To help understand this rather unexpected result,
the light green curves in the bottom panel plot the emission distribution for a subset with deposition peaks separated by $\Delta \mu=0.1$, showing then these generally have much stronger emission away from the apex, due the higher magnetic field strength there.  Thus, even though the black dots show these represent a somewhat lower number in the gaussian distribution, their higher emission leads to an overall peak in the cumulative emission at co-latitudes $|\mu| \approx  0.5 = \sigma_\mu $.

These results have important implications for both the spatial and spectral distribution of radio emission for more realistic models in which the energy deposition has a broader radial distribution, as we discuss next.

%\subsection{Gaussian distribution in radius and co-latitude}
%\subsection{Emission intensity from narrow slice in azimuth}
\subsection{Emission intensity from Gaussian deposition in radius and co-latitude}
\label{sec:intensity}

\begin{figure*}
\includegraphics[width=0.99\textwidth]{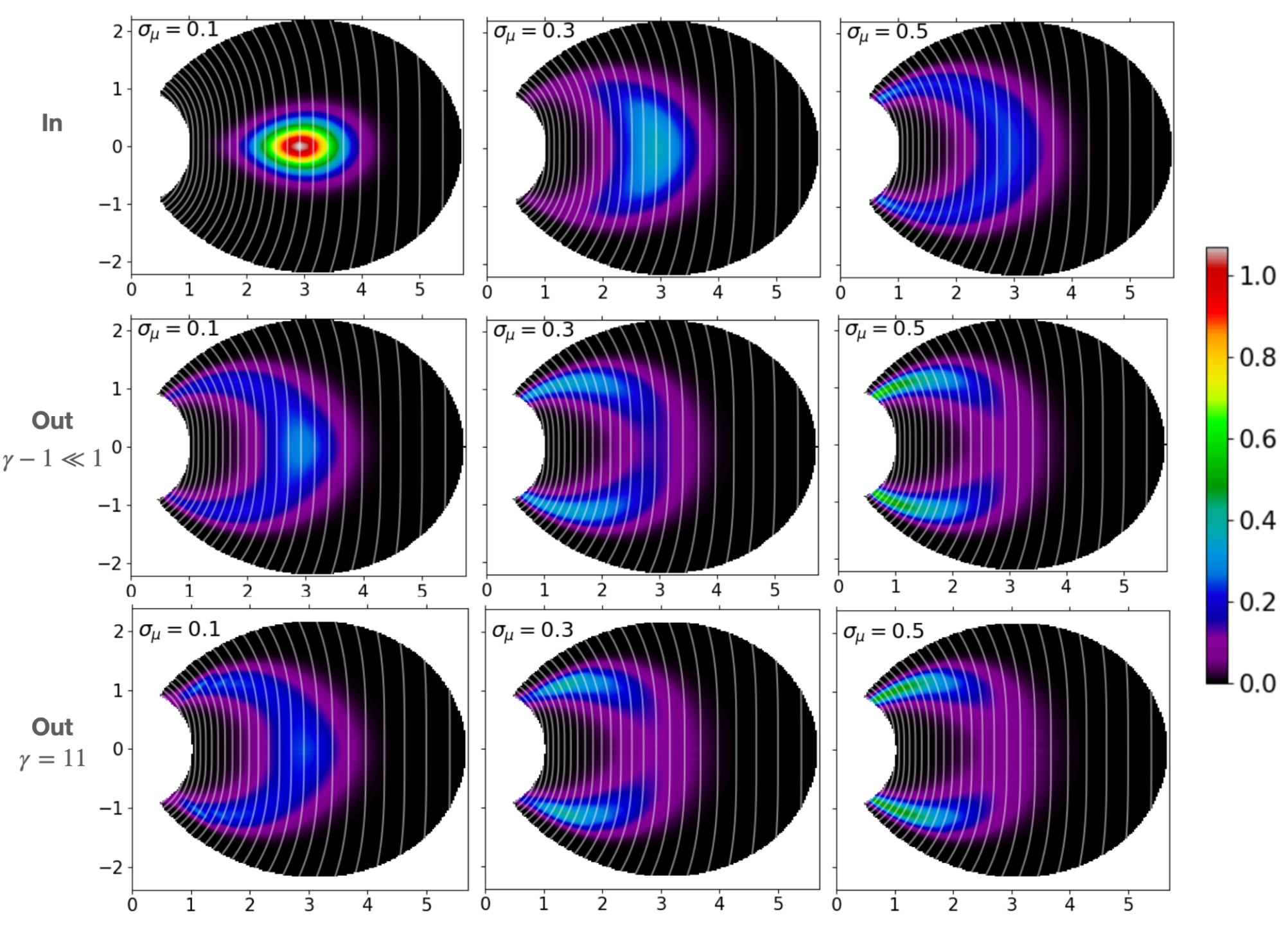}
    \caption{
    Middle row: Spatial variation of emission intensity increment $ dI/d\phi $ (eqn.\ \ref{eq:dIdL}) from a small azimuthal wedge centered on the plane of the sky,
    for the gaussian energy deposition model (eqn.\ \ref{eq:dedrdmu}), assuming $r_p = 3 R_\ast$, $\sigma_r = 0.5 R_\ast$, and the 3  labeled values of $\sigma_\mu = 0.1$, 0.3, and 0.5.
    Top row: Associated intensity for a simpler ``on-the-spot" model, with local emission set to local energy deposition, $\epsilon_{\mu,r_a} = e_{\mu,r_a}$. 
    Bottom row: Same as middle row but for case of relativistic electrons with initial $\gamma=11$.
    Comparison of the top vs.\ middel/bottom rows thus illustrates the differences in energy deposition {\it input} vs.\ emission {\it output} for both relativistic and non-relativisicic electrons.
    In all panels, the total integrated emission is normalized to unity. The white lines represent contours of  magnetic field strength $B$, spaced logarithmically by -0.1 dex from the stellar surface value at the magnetic equator.
    Note that, while these figures do not directly connect to observed radio intensity, they do illustrate how the spatial distribution of energy deposition in the magnetosphere influences the spatial (and also spectral) distribution of resulting radiation by the non-thermal electrons.    
    }
\label{fig:cplot_sigmu}
\label{fig:fig8}
\end{figure*}

Let us now consider a model in which the initial energy deposition has a gaussian spread in both field co-latitude $\mu=\sqrt{1-r/r_a}$ and apex radius $r_a = r/(1-\mu^2)$, centered on a peak radius $r_p$ with radial dispersion $\sigma_{r}$,
\beq
e_{\mu,r_a} =
C_a \exp\left[ -\frac{{(r_a-r_p)}^2}{2\sigma_{r}^2}\right]
~
e_\mu
\, ,
\label{eq:dedrdmu}
\eeq
%This is again normalized 
where $C_a$ is a normalization factor such that 
\begin{align*}
    \int_{R_\ast}^{\infty} \int_{-\mu_\ast}^{+\mu_\ast} e_{\mu,r_a}d\mu\,dr_a&=1
    \, ,
\end{align*}
where $\mu_\ast \equiv \sqrt{1-R_\ast/r_a}$. %is a function of $r_a$.

Given such a distribution in energy deposition $e_{\mu,ra}$, one can, for each $r_a$, use the techniques described above to derive an associated emission distribution $\epsilon_{\mu,r_a}$ along the loop.  
For differential intervals  in loop colatitude $d\mu$ and apex radius $dr_a$, 
%and azimuth $d\phi$, 
the incremental contribution to radio {\it luminosity} is given by
\beq
dL = \epsilon_{\mu,r_a}  d\mu  \,  dr_a 
%\nonumber\\&=& \frac{\epsilon_{\mu,r_a}  d\mu  \, d\phi \, dr}{1-\mu^2}
\, .
\label{eq:dLmrp}
\eeq
Assuming for simplicity that the local emission is isotropic over the full $4\pi$ steradians, the associated increment in {\it intensity} (energy/time/area/solid angle) about an azimuthal interval $d\phi$  centered on the plane of the sky is 
\beq
\frac{dI}{d\phi} = \frac{dL}{4 \pi r \, d\theta \, dr }
= \frac{\epsilon_{\mu,r_a}}
{4 \pi r \, \sqrt{1-\mu^2}}
\, ,
\label{eq:dIdL}
\eeq
where the latter equality uses the relations $d\mu = \sqrt{1-\mu^2} d\theta$ and $dr_a= dr/(1-\mu^2)$.

For a representative, standard case with $r_p=3 R_\ast$ and $\sigma_r = 0.5 R_\ast$, the lower row of Figure \ref{fig:cplot_sigmu} shows the spatial variation of this emitted intensity $dI/d\phi$ for the 3 distinct values of latitudinal dispersion $\sigma_\mu = 0.1$, 0.3, and 0.5.
The white lines show contours of the field strength $B$, spaced logarithmically in increments of $-0.1$ dex from the value at the equatorial surface.

The upper row of Figure \ref{fig:cplot_sigmu} compares the corresponding emission for a simple ``on-the-spot" model, in which the local emission is just set by the local energy deposition,
$\epsilon_{\mu,r_a} = e_{\mu,r_a}$.
Comparison with the middle row shows that the net effect of the electron propagation and mirroring within the closed loops is to shift the emission closer to loop footpoints, where the field strength is higher.
This shift becomes more pronounced with increasing latitudinal dispersion $\sigma_\mu$.

Using the generalized relativistic forms for gyro-cooling derived in Appendix A, the lower row shows a similar spatial distribution for the relativistic case with initial Lorentz factor $\gamma_a = 11$ at the loop apex.
The close similarity to the middle panel shows that the relativistic corrections have only a modest effect on the overall spatial distribution of gyro-cooling emission.
}

The results derived in this subsection  (including Figure \ref{fig:cplot_sigmu}) provides a visualization about the relation between the spatial distribution of energy deposited and that of the energy radiated by the non-thermal electrons. In their current forms, however, these results cannot be compared with observations, which will require a more complete treatment, including absorption effects in the magnetosphere.

\if{false}
\subsection{Associated flux spectrum}

Using the known dipole dependence of  $B(r,\theta)$ to bin the differential luminosity from eqn.\ (\ref{eq:dLmrp}) into intervals of field strength $dL/dB$, one can derive a predicted flux spectrum in $B$,
\beq
F_B \equiv \frac{dL/dB}{L}
\, ,
\label{eq:fB}
\eeq
where $L$ is the total luminosity integrated from the stellar radius outward \footnote{For our assumed simplification of isotropic emission, the axisymmetry of $B$ means that analogous wedges out of the plane of the sky also contribute the same to $L$ and $dL/dB$,  thus increasing both by a factor $2 \pi/d\phi$ to account for the full range of $\phi$, with no net effect on the spectrum $F_B$.}.
Since for our assumed case of sub-relativistic electron energies, the gyro-frequency $\nu_g \sim B$, this also gives a predicted fundamental mode spectrum in observed radio frequency, $F_\nu \sim F_B$.

\if{false}
\begin{figure}
\includegraphics[width=0.49\textwidth]{fractional_flux_spectra_linear.png}
\caption{For the 3 cases of latitudinal dispersion $\sigma_\mu = 0.1$, 0.3, and 0.5, the solid curves show the relative flux spectrum $F_B$ versus magnetic field strength, normalized by the field $B(r_p)$ at the peak deposition radius $r_p = 3 R_\ast$.
For comparison, the dashed curves show the corresponding field distribution for the direct ``on-the-spot" model with $\epsilon_{\mu,r_a}=e_{\mu,r_a}$.}
\label{fig:fB}
\label{fig:fig9}
\end{figure}
\fi

For the 3 cases of latitudinal dispersion, the solid curves of Figure \ref{fig:fB} plot $F_B$ versus $\log(B/B(r_p))$, where $r_p = 3 R_\ast$ is the radius of peak energy deposition, as defined in eqn.\ (\ref{eq:dedrdmu}). For $\sigma_\mu=0.1$, the concentration of emission around this peak radius gives $F_B$ an associated concentration about $B(r_p)$, with a spread that scales with the radial dispersion $\sigma_r$.
But for the higher latitudinal dispersion cases $\sigma_\mu = 0.3$ and 0.5, the peak shifts to slightly higher field strengths, with a increasingly more pronounced tail at higher $B$. 

For a simple model of emission at the fundamental gyro-frequency,
%$\nu_g \sim B$,
\beq
\nu_g \equiv \frac{eB}{m_e c} = 0.28\, {\rm GHz} \, \frac{B}{100 \rm G}
\, ,
\label{eq:nug}
\eeq
these can also represent predictions of observable frequency spectra, with the simple replacements $B/B(r_p) \rightarrow \nu/\nu_g(r_p)$ and $F_B \rightarrow F_\nu$.
More realistically, emission occurs as superpositions of higher-order harmonics of $\nu_g$ \citep[e.g.][]{ramaty1969}, but we leave implementation of this to future work.
\fi

\subsection{Cooling by collisions with thermal electrons}

%\subsection{General scalings}

In addition to the gyro-cooling examined here, non-thermal electrons can also cool by the energy exchange from coulomb collision with an ambient population of thermal electrons of much lower energy
\citep[see][e.g., their section 8.4.3]{2009LNP...778..269G}.
For potentially relativistic electrons with Lorentz factor $\gamma$ and thus  kinetic energy E=$(\gamma -1)m_ec^2$, the coulomb collision cross section takes the form
\beq
\sigma_{c} = 
 \frac{4\pi e^4}{m_e^2 c^4} \,\frac{\ln \Lambda}{
 (\gamma-1)^2
} 
= \frac{3}{2} 
\frac{\ln \Lambda \,
 \sigma_T}{(\gamma-1)^2}
\approx \frac{2 \times 10^{-23} {\rm cm}^2}{ (\gamma-1)^2
} 
\, ,
\label{eq:sigc}
\eeq
where the last evaluation assumes a characteristic value $\ln \Lambda \approx 20$ for the coulomb logarithm (which accounts for the cumulative effect of many small-angle scatterings).
Since each associated collision results in the effective loss of the kinetic energy of the non-thermal particle, the associated cooling time for collisions with thermal electrons of number density $n_e$ is 
\citep[][see their eqn. 1]{1981ApJ...251..781L}
%(LEACH AND PETROSIAN 1981)
\beq
t_c 
%\equiv \frac{E}{dE/dt}
= \frac{\beta}{n_e \sigma_c  c (\gamma-1)}
= 1.7 \times 10^{12} \, {\rm s} \, \frac{\beta (\gamma-1)}{ \, n_e} \, .
\label{eq:tcdef}
\eeq
Comparison with eqn. (\ref{eq:teloss}) shows that the ratio of coulomb to gyro-synchrotron cooling (assuming an isotropic pitch-angle distribution with $\left < \sin^2 \alpha \right > = 2/3$) is given by

\beq
\frac{t_c}{t_e} \approx 2.2 \times
10^9 \, \frac{(\gamma^2 -1) \beta B_{kG}^2}{ n_e} \, ,
\label{eq:tcbte}
\eeq
where $B_{kG} \equiv B/kG$ and $n_e$ is in ${\rm cm}^{-3}$.
Setting the left side to unity, we can solve for the critical density above which
cooling by coulomb collisions will dominate over gyro emission,
\beq
n_{e1} \approx 2.2 \times
10^9 \, (\gamma^2 -1) \beta B_{kG}^2 
\, .
\eeq

For highly relativistic electrons with $\gamma \gg 1$ and $\beta \rightarrow 1$, this critical density increases with $\gamma^2$, implying that in regions with strong kG fields, coulomb cooling will only be important in regions with quite a high density of thermal electrons.

In contrast, for non-relativistic electrons with $\beta < 1$, we find $n_{e1} \sim \beta^3$, implying that cooling by coulomb collisions will dominate over gyro-emission for even modest densities. For example, for the mildly sub-relativistic case $\beta = 1/3$, we find $n_{e1} = 7.2 \times 10^7 \,  {\rm cm}^{-3} \, B_{kG}^2$.

The overall implication is that the incoherent radio emission from magnetic hot stars most likely arises from relativistic electrons for which gyro-emission is relatively unaffected by collisional losses with thermal electrons.
Further quantifying this will depend on the detailed models of the thermal electron density and its spatial distribution. We are currently examining such cooling effects in the context of hot-stars with centrifugal magnetospheres, and will report results in a follow up paper.

\if{false}
\subsubsection{Magnetized stellar wind outflow}

A simple example is an outflowing, magnetized wind, for which the flow tube area scales inversely with field strength, $A \sim A_\ast B_\ast/B$. For mass loss rate ${\dot M}$, local flow speed $V$, and  mean mass per electron $\mu_e$ ($\approx 1.35 m_p$), the electron density scales as 
\beq
n_e = \frac{{\dot M}}{4 \pi R^2
\mu_e V_\infty} 
\frac{b}{w}
= 4.6 \times 10^7 {\rm cm}^{-3} 
\, 
\frac{{\dot M}_{-9}}
{R_{10}^2 V_8} \, \frac{b}{w}
%\frac{B}{B_\ast} 
\, ,
\label{eq:nebwind}
\eeq
where $V_\infty$ is the wind terminal speed,
$w \equiv V/V_\infty$,
$V_8 \equiv V_\infty/(10^8$cm/s),
$b \equiv B/B_\ast$,
$R_{10} \equiv R/10 R_\odot$,
and
${\dot M}_{-9} \equiv {\dot M}/(10^{-9} M_\odot$/yr).
Application of (\ref{eq:nebwind}) into (\ref{eq:tcbte}) gives
\beq
\frac{t_c}{t_e} \approx 50 \, \beta \, (\gamma^2 -1) %\frac{( B/kG)^2 (R/R_\odot)^2  (V/1000 \, km/s)}{{\dot M}/(10^{-8} M_\odot/yr)}
\frac{ B_{kG}^2 R_{10}^2  V_8}{{\dot M}_{-9} }
\, b \, w \, ,
\label{eq:tcbteval}
\eeq
where 
$B_{kG} \equiv B_\ast$/kG.
For velocity variation $w \approx 1-R/r$ and dipole field variation $b \sim (R/r)^3$, the product $b w$ has a peak value of $0.11$ at $r=(4/3) R$, and is smaller at both lower and higher radii. The upshot is that the coulomb cooling time can become comparable or even shorter than the gyro-cooling time in such magnetized winds, at least for electrons that are not too highly relativistic, i.e. with $\gamma \gtrsim 1$.
\fi

\if{false}
{\color{blue}

\subsection{Cooling in the dense CM layer}
To proceed, we need to specify a model for the density distribution of thermal electrons.
For the centrifugal magnetospheres of hot stars, the strongest collisional cooling in the extended CM should be in the dense, compressed layer near the tops of magnetic loops.
%In the wind-fed centrifugal magnetospheres of magnetic B-stars, the outflowing wind is effectively trapped in the region between the Kepler radius $R_{\rm K}$ and the  Alfv\'{e}n radius $R_{\rm A}$. 
The CBO analysis by \citet[][see their eqn.\ 7]{owocki2020} provides a scaling for the radial variation of mass column density in this CM layer.
Through the mean mass per electron $\mu_e = 1.16 m_p$, this can be used to derive an associated electron column density of this thin, dense CM layer,
\beqa
N_e &=& 0.3 
\frac{B_{\rm K}^2}{4 \pi \mu_e g_{{\rm K}}} \, 
\left ( \frac{r_a}{R_{\rm K}} \right )^{-p} 
\nonumber
\\
&=& 4.85 \times 10^{22} {\rm cm}^{-2} \, \frac{B_{kG}^2}{g_4} 
\left ( \frac{R_{\rm K}}{R} \right )^{-4}
\left ( \frac{r_a}{R_{\rm K}} \right )^{-5} 
\, ,
\label{eq:NeCBO}
\eeqa
where $B_{\rm K} $and $g_{\rm K}$ are the magnetic field strength and stellar gravity at the Kepler radius, and the power index allows for generalizations from the original $p=6$ assumed in eqn.\ 7 of \citet{owocki2020}.

Following \citet{Berry2022} and \citet{ud2023}, the latter equality
instead takes $p=5$, and provides numerical scalings in terms of the scaled surface gravity $g_4 \equiv g_\ast/(10^4 \,$cm/s$^2$), and the polar surface field in kG, $B_{kG}\equiv B_{p\ast
 }/$kG.
 For a star with critical rotation fraction $W$ at its equatorial surface, $R_{\rm K}/R = W^{-2/3}$.

 For high-energy electrons passing through this CM layer, the associated collision depth is thus
 \beq
 \tau_c = N_e \sigma_c =
 0.96 \, \frac{B_{kG}^2}{g_4 (\gamma-1)^2} 
%\left ( \frac{R_{\rm K}}{R} \right )^{-4}
(2W)^{8/3}
\left ( \frac{r_a}{R_{\rm K}} \right )^{-5} 
\, .
 \label{eq:tauc}
\eeq
For each pass through the CM layer, the fractional energy loss is $1-e^{-\tau_c}$.
For sub-relativistic electrons in loops with apexes not too far above the Kepler radius, we infer $\tau_c > 1$, implying coulomb collisions in the CM layer will dissipate the  electron energy in the first few CM crossings, before there is much mirroring and associated gyro-emission.

The upshot is that such coulomb cooling in the dense CM layer should effectively {\it quench} the radio emission from sub-relativistic electrons near $R_K$.

But for relativistic electrons sufficiently above $R_K$, we find $1-e^{-\tau_c} \approx \tau_c \ll 1$,
implying the energy loss will again occur over many mirror cycles.
Let us thus now estimate the scalings for relative loss to gyro-cooling vs.\ coulomb cooling when accumulated over these cycles.

\subsection{Reduction of gyro-emission}

The constant $k$ derived in eqn.\ (\ref{eq:kdef})
represents the competing loss to gyro-cooling during each mirror cycle.
Casting this in terms of these Kepler-radius parameter scalings, we have 
\beqa
~~~~~~~~~~k &=&  
\frac{B_{\rm K}^2 R_K}{2 \pi m_e c^2 \beta_a} \, 
\left ( \frac{r_a}{R_{\rm K}} \right )^{-5} 
\nonumber
\\
&=&  0.032 \, \frac{ B_{kG}^2 R_{12} }{\beta_a} (2W)^{10/3}
\left ( \frac{r_a}{R_{\rm K}} \right )^{-5} 
\, ,
\label{eq:kK}
\eeqa
where $R_{12} \equiv R/(10^{12}$cm) and $\beta_a \equiv v_a/c$.

Comparison of eqn.\ (\ref{eq:kK}) with (\ref{eq:tauc}) shows that, quite remarkably,  both gyro-cooling along magnetic loops and coulomb cooling in the CM layer have the {\it same} dependence on the magnitude and radial variation of magnetic field, implying then that their ratio is a {\it global} parameter that is  fully {\it independent} of the field\footnote{Note that $g_4 R_{12} = V_8^2$, where $V_8 \equiv V_{orb}/(10^8$\,cm/s) is a scaled form for the orbital speed $V_{orb} = \sqrt{GM/R}$ at the stellar surface.},
\beq
\boxed{
\frac{k}{\tau_c} \approx 0.0033 \, \frac{(\gamma -1)^2}{\beta_a} \, 
(2W)^{2/3}\, g_4 R_{12}
}
\, .
\label{eq:kbtauc}
\eeq
We can use this to define an estimated cooling reduction factor,
\beq
f_c \equiv \frac{k/\tau_c}{1+k/\tau_c}
\, .
\label{eq:fcdef}
\eeq
As long as $\tau_c \ll 1$ -- which applies for relativistic electrons not too near the Kepler radius --,
$f_c$ characterizes the
 overall fractional reduction of gyro-synchrotron radio emission that results from the cumulative energy loss by coulomb collisions from many mirror passings through the dense CM layer.
%Quite remarkably, because coulomb cooling in the CM layer and gyro-cooling along magnetic loops both scale with the magnetic field strength at the loop apex, their relative importance is largely {\it independent} of both the field strength and its decline with radius

Eqn. (\ref{eq:kbtauc} shows that
this reduction can be quite substantial, even for mildly relativistic electrons.
For example, for $\gamma=2$ and standard stellar parameters ($2W=g_4=R_{12}=1$), $f_c = 0.003$, implying only $0.3\%$ of the initial electron energy ends up as gyro-synchrotron radio emission.
For the moderately relativistic example $\gamma=11$ considered above, we find $f_c \approx 0.25$,
while for the highly relativistic case $\gamma=101$, $f_c \approx 0.97$.
}
\fi
\if{false}
But importantly, in all these relativistic cases for which $\tau_c \ll 1$, the overall energy loss again occurs through multiple mirror cycles, much as occurs when $k \ll 1$ in the above analysis without coulomb cooling.
With such additional cooling, the net number of cycles is set by $\max(k,\tau_c)$; but because any coulomb losses are confined to a narrow CM layer, the relative spatial distribution of any residual gyro-emission should not be much affected.

{\color{black}
NOTE to Barnali:  I'm now rethinking this argument, because it fails to account for the fact that different pitch angles will have different rates of crossing the CM layer.
Thus working out the effect on spatial distribution probably requires more detailed calculation, with a cooling rate that depends on the CM crossing rate.
Because the ratio $k/\tau_c$ is a global constant, this would only have to be done for a single model loop for any given  value of $k/\tau_c$, which however depends on the assumed energy through $\gamma$.  ;-(
Let's discuss this.
}

Thus, for the above sample relativistic case $\gamma = 11$, the relative spatial {\it distribution} of emission shown in the lower row of figure \ref{fig:fig8} should still apply for models with coulomb cooling in the CM layer, with just the overall radio emission reduced by the factor $f_c$ given by eqns.\ (\ref{eq:kbtauc} and (\ref{eq:fcdef}.

Finally, the analysis here is based on scalings inferred 
\citep[e.g., ][]{owocki2020} for the case of a dipole field aligned with the stellar rotation axis.
For tilted dipoles, recent 3D MHD simulations indicate a more complex CM geometry \citep{ud2023}, and so future work should explore how this might affect the relative importance of coulomb cooling.
}
\fi
\section{Summary and Future Work}

\subsection{Result summary}

The incoherent, circularly polarized radio emission observed from magnetic massive stars is understood to arise from gyro-synchrotron emission by energetic electrons trapped within their  magnetospheres.
By accounting for the associated ``gyro-cooling'' loss of energy, this paper analyzes the spatial distribution of this radio emission for a simple model in which the electrons trapped within closed (assumed dipole) magnetic loops gradually cool as they repeatedly mirror across the loop.
Some key results are:
\begin{itemize}
%\item 
%In contrast to the decade-long gyro-cooling time for solar and planetary magnetospheres, the much greater ($>$100\,G) field strengths of magnetic massive stars yields cooling times of order a day or less.
\item
Because the gyro-cooling time in hot-star magnetospheres is of order a day,
explaining their observed persistent quasi-steady radio emission requires a persistent electron acceleration mechanism, perhaps from  magnetic reconnection driven by repeated, small-scale centrifugal breakout (CBO) events in these rapidly rotating magnetospheres
\citep{owocki2022}.
%We find that 
\item
So long as the ratio of the advective propagation time to cooling time is small, $k \equiv t_a/t_c \ll 1$, the spatial distribution of cumulative emission is insensitive to the specific value of this ratio.
\item
 For various assumptions for the spatial location of the {\it deposition} of  energetic electrons  centered about the apex of closed magnetic loops, the associated radio {\it emission} tends to be spread down the loop where the field is stronger (see Figure \ref{fig:fig8}).
 %\item
 %With current observing facilities, it is not possible to directly observe the spatial distribution of emitted energy in the stellar magnetospheres. However, upcoming facilities such as the ngVLA \citep{mckinnon2019} will offer unprecedented sensitivity and angular resolution from mid to high radio frequencies (a few GHz to $\sim 100$ GHz) that will enable to obtain resolved radio images of the magnetospheres of at least a subset of hot magnetic stars. By combining such images acquired over a wide range of radio frequencies, one will be able to directly compare them with the prediction of our simple model, and infer characteristics of the underlying energy deposition process.
 %\item
 The formalism here thus effectively represents an energy transport model between the deposition of energetic electrons to their ultimate gyro-synchrotron radio emission. 
 
 \item 
 Generalization to the case of relativistic electrons introduces a formal, explicit dependence of gyro-cooling on the initial energy, but the overall effect on the resulting spatial distribution of emission is quite modest.
 
 \item
 Cooling from coulomb collisions with thermal electrons in the dense CM layer should effectively quench any gyro-emission from sub-relativistic electrons, but have less effect on relativistic electrons.
 
{\if{false}
\item
In the simple case for which the emission frequency is dominated by the fundamental mode at the cyclotron frequency, this spatial transport model has direct implications for the observed radio spectrum. 
\item
Specifically, we show here that it leads to a predicted radio spectrum that extends upward from the frequency associated with the apex field near the peak deposition radius, to higher frequencies associated with the stronger field near the loop footpoints. Figure \ref{fig:fig9} provides specific results for 3 sample models for the  energy deposition.
\fi
}
\end{itemize}

\subsection{Future extensions}
While the work here thus provides an initial basis for modelling the spatial and spectral distribution of radio emission from these massive-star magnetospheres, it is grounded in several idealized assumptions and simplifications that should be examined and relaxed in future work.
\begin{itemize} 

\item 
To translate results for spatial distribution of emission within magnetic loops into observable spectra in frequency,
future work should consider a  realistic scenario of multiple harmonic emission and absorption in the stellar magnetosphere.

\item 
This should include the effect of cooling of the energetic electrons through coulomb collision with thermal electrons associated with material trapped within the centrifugal magnetosphere.

\item In contrast to the static, dipole field assumed here, even closed loops in centrifugal magnetospheres are likely to be dynamically distorted, due to the stretching by the centrifugal force from the trapped material and the associated breakout events. Future work should examine how this affects electron mirroring and the associated emission.
\item As found in planetary magnetospheres, such variations can excite MHD waves and even cascade to magnetic turbulence, leading to pitch-angle scattering of electrons \citep[e.g.][]{summers_mace_hellberg_2005,kim2018} into a loss cone that allows interaction with the underlying planet or star. The associated energy loss can compete with gyro-cooling, and so lower the overall radio emission.
\item Both loss mechanisms lead to formation of anisotropic electron distribution. Future work should examine how this affects the associated electron cyclotron maser emission (ECME) seen from many such magnetic stars \citep[e.g.][]{2000AA...362..281T,das2022}.
\item Instead of the simplified assumption here of isotropic emission, future work should take into account the distinctive angular phase function for gyro-synchrotron emission.
\item  Within the CBO-driven reconnection paradigm, there is a need for detailed modeling, e.g. using particle-in-cell (PIC) codes \citep{Lapenta2012,Germaschewski2016}, of the electron electron acceleration from reconnection, with a focus on the spatial deposition of non-thermal electrons.
%and their associated pitch-angle distribution.
\item Does such reconnection yield non-gyrotropic electron deposition, and if so, does it favor field-aligned or orthogonal pitch angles? 
This is crucial for mirroring and the spatial distribution of radio emission.
\item Finally, to test and constrain the basic gyro-cooling model and any extensions, there is a need for more extended observational programs on radio spectra and  their level of variability on a dynamical timescale of order a day.
While current radio observatories cannot spatially resolve the emission from hot-star magnetospheres, upcoming facilities such as the ngVLA \citep{mckinnon2019} will have the  sensitivity and angular resolution to do so for at least a subset of hot magnetic stars; this will  allow direct tests of predictions from our simple model, and constrain characteristics of the underlying energy deposition processes.

\end{itemize}

Overall, despite the relatively idealized nature of its basic assumptions, the gyro-cooling analysis presented here forms a good initial basis for modelling the observed incoherent radio emission from these magnetic massive stars.

\if{false}
{\color{black}

\section{Equations with coulomb cooling}

Using eqn.\ (\ref{eq:tcdef}), the apex-time-scaled rate for local coulomb cooling can be written
\beqa
~~~~~~~-{\dot e}_c = e \frac{t_a}{t_c} &=& e \frac{n_e \sigma_c (\gamma -1)}{\beta} \, \frac{r_a}{v_a}
\nonumber
\\
&=& \frac{\tau_c (\gamma-1)^2}{\beta \beta_a (\gamma_a-1)} \, \frac{{\rm e}^{-(\mu/\mu_H)^2}}{\sqrt{\pi} \mu_H}
\nonumber
%\\ &=& \frac{\tau_c (\gamma_a-1)}{\beta \beta_a} \, \frac{{\rm e}^{-(\mu/\mu_H)^2}}{\sqrt{\pi} \mu_H}
\, ,
\eeqa
where the latter expressions uses the gaussian density stratification of the CM layer over a scale height
given by \citep[see][their eqn.\ 4 ]{Owocki2018}
\beq
H = \frac{\sqrt{2} c_s/\Omega}{\sqrt{3-2(R_K/r_a)^3}} 
\, ,
\eeq
with $\Omega$ the stellar rotation frequency.
The corresponding scale height in co-latitude $\mu$ is
\beq
\mu_H \equiv  \frac{H}{r_a} = \frac{0.057}{2W \sqrt{g_4 R_{12}}\sqrt{3-2(R_K/r_a)^3}}
\, \frac{R}{r_a}
\, .
\label{eq:muHdef}
\eeq
where the latter evaluation assumes an isothermal sound speed $c_s = 20$\,km/s.

Using eqn.\ (\ref{eq:kbtauc}) to eliminate $\tau_c$ in favor of the gyro-cooling parameter $k$, we find
for standard stellar parameters $2W=g_4=R_{12}=1$,
\beq
\boxed{
{\dot e}_c = - \, \frac{300 k}{\beta (\gamma_a-1)} \, \frac{{\rm e}^{-(\mu/\mu_H)^2}}{\sqrt{\pi}\mu_H}
}
\, ,
\label{eq:edotcstd}
\eeq
with 
\beq
\mu_H = \frac{0.057}{\sqrt{3-8(R/r_a)^3}}
\, \frac{R}{r_a}
\, ~ ; ~ r_a \ge R_K = 1.59 R
\eeq
For general parameter values, ${\dot e}_c$ is modified by a factor $1/((2W)^{2/3}  g_4 R_{12})$ from the value in eqn.\ (\ref{eq:edotcstd}).

By comparison, eqn.\c(\ref{eq:energy_eqn}) gives the relativistic form for the competing gyro-cooling emission term. Noting that the relativistic correction factor in curly brackets just becomes $\gamma -1$, their ratio can be characterized by
\beq
\frac{\left < {\dot e}_e \right >}{\left < {\dot e}_c\right >} = 0.0033\,  \beta (\gamma_a-1)(\gamma+1)\left <pb^3 \right >
\, ,
\eeq
where the angle brackets denote the cumulative average from the apex $\mu=0$ to the mirror point $\mu=\mu_m$.

This ratio has a similar scaling to that derived for $k/\tau_c$ in eqn.\ (\ref{eq:kbtauc}), but the factor $\left < pb^3 \right >$ can become large for field aligned pitch angles, for which the mirror co-latitude $\mu_m \lesssim 1$, with thus $b \sim 1/(1-\mu_m^2)^3 \gg 1$  implying strong gyro-emission near this low-lying mirror point.

This indicates coulomb cooling should have comparatively greater effect for more perpendicular pitch angles that mirror closer to the loop apex.
The overall effect should be to shift the spatial distribution of any residual gyro-emission to further down the loop footpoints.

A good test case would be to implement this for the standard relativistic case $\gamma_a=11$.
}
\fi

\section*{Acknowledgements}
We thank the referee for their constructive comments that helped us to improve the manuscript significantly.
BD acknowledges support from the Bartol Research Institute. The contributions by SPO are supported in part by the National Aeronautics and Space Administration under Grant No. 80NSSC22K0628 issued through the Astrophysics Theory Program.

\section*{Data availability}
This is a theoretical work, and does not use any observational data.

\bibliography{main}{}

\begin{thebibliography}{}
\makeatletter
\relax
\def\mn@urlcharsother{\let\do\@makeother \do\$\do\&\do\#\do\^\do\_\do\%\do\~}
\def\mn@doi{\begingroup\mn@urlcharsother \@ifnextchar [ {\mn@doi@}
  {\mn@doi@[]}}
\def\mn@doi@[#1]#2{\def\@tempa{#1}\ifx\@tempa\@empty \href
  {http://dx.doi.org/#2} {doi:#2}\else \href {http://dx.doi.org/#2} {#1}\fi
  \endgroup}
\def\mn@eprint#1#2{\mn@eprint@#1:#2::\@nil}
\def\mn@eprint@arXiv#1{\href {http://arxiv.org/abs/#1} {{\tt arXiv:#1}}}
\def\mn@eprint@dblp#1{\href {http://dblp.uni-trier.de/rec/bibtex/#1.xml}
  {dblp:#1}}
\def\mn@eprint@#1:#2:#3:#4\@nil{\def\@tempa {#1}\def\@tempb {#2}\def\@tempc
  {#3}\ifx \@tempc \@empty \let \@tempc \@tempb \let \@tempb \@tempa \fi \ifx
  \@tempb \@empty \def\@tempb {arXiv}\fi \@ifundefined
  {mn@eprint@\@tempb}{\@tempb:\@tempc}{\expandafter \expandafter \csname
  mn@eprint@\@tempb\endcsname \expandafter{\@tempc}}}

\bibitem[\protect\citeauthoryear{{Andre}, {Montmerle}, {Feigelson}, {Stine}  \&
  {Klein}}{{Andre} et~al.}{1988}]{andre1988}
{Andre} P.,  {Montmerle} T.,  {Feigelson} E.~D.,  {Stine} P.~C.,   {Klein}
  K.-L.,  1988, \mn@doi [\apj] {10.1086/166979}, \href
  {https://ui.adsabs.harvard.edu/abs/1988ApJ...335..940A} {335, 940}

\bibitem[\protect\citeauthoryear{{Auri{\`e}re} et~al.,}{{Auri{\`e}re}
  et~al.}{2007}]{2007AA...475.1053A}
{Auri{\`e}re} M.,  et~al., 2007, \mn@doi [\aap] {10.1051/0004-6361:20078189},
  \href {http://adsabs.harvard.edu/abs/2007A26A...475.1053A} {475, 1053}

\bibitem[\protect\citeauthoryear{{Condon} \& {Ransom}}{{Condon} \&
  {Ransom}}{2016}]{ConRan16}
{Condon} J.~J.,  {Ransom} S.~M.,  2016, {Essential Radio Astronomy}.
\url {https://www.cv.nrao.edu/~sransom/web/Ch5.html}

\bibitem[\protect\citeauthoryear{{Das} \& {Chandra}}{{Das} \&
  {Chandra}}{2021}]{das2021}
{Das} B.,  {Chandra} P.,  2021, \mn@doi [\apj] {10.3847/1538-4357/ac1075},
  \href {https://ui.adsabs.harvard.edu/abs/2021ApJ...921....9D} {921, 9}

\bibitem[\protect\citeauthoryear{{Das} et~al.,}{{Das} et~al.}{2022}]{das2022}
{Das} B.,  et~al., 2022, \mn@doi [\apj] {10.3847/1538-4357/ac2576}, \href
  {https://ui.adsabs.harvard.edu/abs/2022ApJ...925..125D} {925, 125}

\bibitem[\protect\citeauthoryear{{Drake}, {Abbott}, {Bastian}, {Bieging},
  {Churchwell}, {Dulk}  \& {Linsky}}{{Drake}
  et~al.}{1987}]{1987ApJ...322..902D}
{Drake} S.~A.,  {Abbott} D.~C.,  {Bastian} T.~S.,  {Bieging} J.~H.,
  {Churchwell} E.,  {Dulk} G.,   {Linsky} J.~L.,  1987, \mn@doi [\apj]
  {10.1086/165784}, \href {http://adsabs.harvard.edu/abs/1987ApJ...322..902D}
  {322, 902}

\bibitem[\protect\citeauthoryear{{Germaschewski}, {Fox}, {Abbott}, {Ahmadi},
  {Maynard}, {Wang}, {Ruhl}  \& {Bhattacharjee}}{{Germaschewski}
  et~al.}{2016}]{Germaschewski2016}
{Germaschewski} K.,  {Fox} W.,  {Abbott} S.,  {Ahmadi} N.,  {Maynard} K.,
  {Wang} L.,  {Ruhl} H.,   {Bhattacharjee} A.,  2016, \mn@doi [Journal of
  Computational Physics] {10.1016/j.jcp.2016.05.013}, \href
  {https://ui.adsabs.harvard.edu/abs/2016JCoPh.318..305G} {318, 305}

\bibitem[\protect\citeauthoryear{{Grunhut} et~al.,}{{Grunhut}
  et~al.}{2017}]{2017MNRAS.465.2432G}
{Grunhut} J.~H.,  et~al., 2017, \mn@doi [\mnras] {10.1093/mnras/stw2743}, \href
  {http://adsabs.harvard.edu/abs/2017MNRAS.465.2432G} {465, 2432}

\bibitem[\protect\citeauthoryear{{G{\"u}del}}{{G{\"u}del}}{2009}]{2009LNP...778..269G}
{G{\"u}del} M.,  2009, in {Cargill} P.,  {Vlahos} L.,  eds, , Vol.~778,
  Turbulence in Space Plasmas.
p.~269, \mn@doi{10.1007/978-3-642-00210-6_8}

\bibitem[\protect\citeauthoryear{{Kim}, {Kim}  \& {Kwon}}{{Kim}
  et~al.}{2018}]{kim2018}
{Kim} K.-H.,  {Kim} G.-J.,   {Kwon} H.-J.,  2018, \mn@doi [Earth, Planets and
  Space] {10.1186/s40623-018-0947-9}, \href
  {https://ui.adsabs.harvard.edu/abs/2018EP&S...70..174K} {70, 174}

\bibitem[\protect\citeauthoryear{{Kochukhov}, {Shultz}  \&
  {Neiner}}{{Kochukhov} et~al.}{2019}]{2019A&A...621A..47K}
{Kochukhov} O.,  {Shultz} M.,   {Neiner} C.,  2019, \mn@doi [\aap]
  {10.1051/0004-6361/201834279}, \href
  {http://adsabs.harvard.edu/abs/2019A%26A...621A..47K} {621, A47}

\bibitem[\protect\citeauthoryear{{Lapenta}}{{Lapenta}}{2012}]{Lapenta2012}
{Lapenta} G.,  2012, \mn@doi [Journal of Computational Physics]
  {10.1016/j.jcp.2011.03.035}, \href
  {https://ui.adsabs.harvard.edu/abs/2012JCoPh.231..795L} {231, 795}

\bibitem[\protect\citeauthoryear{{Leach} \& {Petrosian}}{{Leach} \&
  {Petrosian}}{1981}]{1981ApJ...251..781L}
{Leach} J.,  {Petrosian} V.,  1981, \mn@doi [\apj] {10.1086/159521}, \href
  {https://ui.adsabs.harvard.edu/abs/1981ApJ...251..781L} {251, 781}

\bibitem[\protect\citeauthoryear{{Leto}, {Trigilio}, {Buemi}, {Umana}  \&
  {Leone}}{{Leto} et~al.}{2006}]{2006A&A...458..831L}
{Leto} P.,  {Trigilio} C.,  {Buemi} C.~S.,  {Umana} G.,   {Leone} F.,  2006,
  \mn@doi [\aap] {10.1051/0004-6361:20054511}, \href
  {http://adsabs.harvard.edu/abs/2006A%26A...458..831L} {458, 831}

\bibitem[\protect\citeauthoryear{{Leto}, {Trigilio}, {Buemi}, {Leone}  \&
  {Umana}}{{Leto} et~al.}{2012}]{leto2012}
{Leto} P.,  {Trigilio} C.,  {Buemi} C.~S.,  {Leone} F.,   {Umana} G.,  2012,
  \mn@doi [\mnras] {10.1111/j.1365-2966.2012.20997.x}, \href
  {http://cdsads.u-strasbg.fr/abs/2012MNRAS.423.1766L} {423, 1766}

\bibitem[\protect\citeauthoryear{{Leto} et~al.,}{{Leto}
  et~al.}{2021}]{leto2021}
{Leto} P.,  et~al., 2021, \mn@doi [\mnras] {10.1093/mnras/stab2168}, \href
  {https://ui.adsabs.harvard.edu/abs/2021MNRAS.507.1979L} {507, 1979}

\bibitem[\protect\citeauthoryear{{McKinnon}, {Beasley}, {Murphy}, {Selina},
  {Farnsworth}  \& {Walter}}{{McKinnon} et~al.}{2019}]{mckinnon2019}
{McKinnon} M.,  {Beasley} A.,  {Murphy} E.,  {Selina} R.,  {Farnsworth} R.,
  {Walter} A.,  2019, in Bulletin of the American Astronomical Society. p.~81

\bibitem[\protect\citeauthoryear{{Owocki}, {Shultz}, {ud-Doula}, {Sundqvist},
  {Townsend}  \& {Cranmer}}{{Owocki} et~al.}{2020}]{owocki2020}
{Owocki} S.~P.,  {Shultz} M.~E.,  {ud-Doula} A.,  {Sundqvist} J.~O.,
  {Townsend} R. H.~D.,   {Cranmer} S.~R.,  2020, \mn@doi [\mnras]
  {10.1093/mnras/staa2325}, \href
  {https://ui.adsabs.harvard.edu/abs/2020MNRAS.499.5366O} {499, 5366}

\bibitem[\protect\citeauthoryear{{Owocki}, {Shultz}, {ud-Doula}, {Chandra},
  {Das}  \& {Leto}}{{Owocki} et~al.}{2022}]{owocki2022}
{Owocki} S.~P.,  {Shultz} M.~E.,  {ud-Doula} A.,  {Chandra} P.,  {Das} B.,
  {Leto} P.,  2022, \mn@doi [\mnras] {10.1093/mnras/stac341}, \href
  {https://ui.adsabs.harvard.edu/abs/2022MNRAS.513.1449O} {513, 1449}

\bibitem[\protect\citeauthoryear{{Petit} et~al.,}{{Petit}
  et~al.}{2013}]{petit2013}
{Petit} V.,  et~al., 2013, \mn@doi [\mnras] {10.1093/mnras/sts344}, \href
  {http://adsabs.harvard.edu/abs/2013MNRAS.429..398P} {429, 398}

\bibitem[\protect\citeauthoryear{{Shultz} et~al.,}{{Shultz}
  et~al.}{2019}]{2019MNRAS.482.3950S}
{Shultz} M.,  et~al., 2019, \mn@doi [\mnras] {10.1093/mnras/sty2985}, \href
  {http://adsabs.harvard.edu/abs/2019MNRAS.482.3950S} {482, 3950}

\bibitem[\protect\citeauthoryear{{Shultz} et~al.,}{{Shultz}
  et~al.}{2020}]{shultz2020}
{Shultz} M.~E.,  et~al., 2020, \mn@doi [\mnras] {10.1093/mnras/staa3102}, \href
  {https://ui.adsabs.harvard.edu/abs/2020MNRAS.499.5379S} {499, 5379}

\bibitem[\protect\citeauthoryear{{Shultz} et~al.,}{{Shultz}
  et~al.}{2022}]{shultz2021}
{Shultz} M.~E.,  et~al., 2022, \mn@doi [\mnras] {10.1093/mnras/stac136}, \href
  {https://ui.adsabs.harvard.edu/abs/2022MNRAS.513.1429S} {513, 1429}

\bibitem[\protect\citeauthoryear{{Sikora}, {Wade}, {Power}  \&
  {Neiner}}{{Sikora} et~al.}{2019}]{2019MNRAS.483.2300S}
{Sikora} J.,  {Wade} G.~A.,  {Power} J.,   {Neiner} C.,  2019, \mn@doi [\mnras]
  {10.1093/mnras/sty3105}, \href
  {http://adsabs.harvard.edu/abs/2019MNRAS.483.2300S} {483, 2300}

\bibitem[\protect\citeauthoryear{Summers, Mace  \& Hellberg}{Summers
  et~al.}{2005}]{summers_mace_hellberg_2005}
Summers D.,  Mace R.~L.,   Hellberg M.~A.,  2005, \mn@doi [Journal of Plasma
  Physics] {10.1017/S0022377804003186}, 71, 237–250

\bibitem[\protect\citeauthoryear{{Trigilio}, {Leto}, {Leone}, {Umana}  \&
  {Buemi}}{{Trigilio} et~al.}{2000}]{2000AA...362..281T}
{Trigilio} C.,  {Leto} P.,  {Leone} F.,  {Umana} G.,   {Buemi} C.,  2000, \aap,
  \href {http://adsabs.harvard.edu/abs/2000A%26A...362..281T} {362, 281}

\bibitem[\protect\citeauthoryear{{Trigilio}, {Leto}, {Umana}, {Leone}  \&
  {Buemi}}{{Trigilio} et~al.}{2004}]{2004A&A...418..593T}
{Trigilio} C.,  {Leto} P.,  {Umana} G.,  {Leone} F.,   {Buemi} C.~S.,  2004,
  \mn@doi [\aap] {10.1051/0004-6361:20040060}, \href
  {http://adsabs.harvard.edu/abs/2004A\%26A...418..593T} {418, 593}

\bibitem[\protect\citeauthoryear{Virtanen et~al.,}{Virtanen
  et~al.}{2020}]{2020SciPy-NMeth}
Virtanen P.,  et~al., 2020, \mn@doi [Nature Methods]
  {10.1038/s41592-019-0686-2}, \href {https://rdcu.be/b08Wh} {17, 261}

\bibitem[\protect\citeauthoryear{{ud-Doula}, {Owocki}  \&
  {Townsend}}{{ud-Doula} et~al.}{2009}]{ud2009}
{ud-Doula} A.,  {Owocki} S.~P.,   {Townsend} R.~H.~D.,  2009, \mn@doi [MNRAS]
  {10.1111/j.1365-2966.2008.14134.x}, \href
  {http://adsabs.harvard.edu/abs/2009MNRAS.392.1022U} {392, 1022}

\bibitem[\protect\citeauthoryear{von~der Linden, Fiksel, Peebles, Edwards,
  Willingale, Link, Mastrosimone  \& Chen}{von~der Linden
  et~al.}{2021}]{linden2021}
von~der Linden J.,  Fiksel G.,  Peebles J.,  Edwards M.~R.,  Willingale L.,
  Link A.,  Mastrosimone D.,   Chen H.,  2021, \mn@doi [Physics of Plasmas]
  {10.1063/5.0057582}, 28

\makeatother
\end{thebibliography}

\appendix

\section{Relativistic equations}\label{subsec:relativistic_eqns}
Let us now consider the generalization of the gyro-cooling equations for the relativistic case  of electrons with an initial Lorentz factor $\gamma_a > 1$ at  the loop apex, thus with  associated energy %at the apex (equivalent to energy at $t=0$) 
is $E_a=(\gamma_a-1)mc^2$. 

The corresponding apex-scaled energy along the loop is thus now given by
\begin{align*}
    e&=\frac{E}{E_a}=\frac{\gamma-1}{\gamma_a-1} \, ,
\end{align*}
so that
\begin{align}
    \gamma&=1+(\gamma_a-1)e 
    \, .
    \label{eq:gamma_eqn}
\end{align}
%We also define $p=(e/b)\sin^2\alpha$, where $b=B/B_a$, and $\alpha$ is the pitch angle. 
The relativistic magnetic moment is  given by \citep[e.g.][]{linden2021}:
\begin{align*}
    P_m&=\frac{\gamma m v^2\sin^2\alpha}{2B} \, .
\end{align*}
Substituting $\sin^2\alpha=pb/e$, $B=bB_a$, $v^2=\beta^2c^2=c^2(\gamma^2-1)/\gamma^2$, we get
\begin{align}
    P_m&=\left(\frac{\gamma+1}{2\gamma}\right)\frac{pE_a}{B_a} \, .
    \label{eq:magnetic_moment}
\end{align}

Let us first consider the equation for the time evolution of $\mu=\cos\theta$. By setting $t\equiv t/t_a$, where $t_a=r_a/v_a$, we have
\beq
    \frac{d\mu}{dt}=-\frac{\beta\cos\alpha}{\beta_a\sqrt{1+3\mu^2}}
\, ,
\nonumber
\eeq
which reduces to
%    \nonumber\\
%&= -\mathrm{Sign}%(\cos\alpha) \frac{\sqrt{e-pb}}{\sqrt{1+3\mu^2}}
%\left { 
%\frac{\beta}{\beta_a\sqrt{e}} 
%\right }
%    \end{align}
%\Rightarrow 
\beq
\boxed{
\frac{d\mu}{dt}    =-\mathrm{Sign}(\cos\alpha) \,
\frac{\sqrt{e-pb}}{\sqrt{1+3\mu^2}}
\, \left \{ \frac{\beta}{\beta_a\sqrt{e}} \right \}
} \, ,
\label{eq:relmu}
\eeq
where
\beq
%F(e,\gamma_a) =
    \frac{\beta}{\beta_a\sqrt{e}}
    =\frac{\gamma_a\sqrt{2+(\gamma_a-1)e}}{(1+(\gamma_a-1)e)\sqrt{\gamma_a+1}} \, .
\eeq
Note that this relativistic correction factor reduces to unity when
$\gamma_a \rightarrow 1$.

Let us next consider the relativistic form of the energy equation,
\begin{align*}
    \frac{dE}{dt}&=-\frac{\sigma_\mathrm{T}c}{4\pi}\beta^2\gamma^2B^2\sin^2\alpha
    \, .
\end{align*}
Again setting $t \rightarrow t/t_a$, $E=eE_a$, $B=bB_a$ and $\sin^2\alpha=pb/e$,
we find the relativistic generalization of the scaled energy equation (\ref{eq:dedt} takes the form,
%to get:
%\begin{align*}
%    \frac{de}{dt}&=\left(\frac{\sigma_\mathrm{T}B_a^2r_a}{2\pi m cv_a}\right)\left(\frac{\gamma+1}{2}\right)pb^3
%\end{align*}

%we get:
\beq
\boxed{
\frac{de}{dt}=-k \, pb^3 \left\{1+ \frac{e(\gamma_a -1)}{2}\right\} } \, ,
\label{eq:energy_eqn}
\eeq
where the cooling constant $k$ is defined by eqn.\ (\ref{eq:kdef}).

Finally, we consider the equation for the time evolution of magnetic moment,
\begin{align*}
    \frac{d}{dt}(\gamma m v \cos\alpha)&=-\frac{P_m}{v\cos\alpha}\frac{dB}{dt}
\end{align*}
Using eqns.\ \ref{eq:gamma_eqn}, \ref{eq:magnetic_moment} and \ref{eq:energy_eqn}, we can cast this into the scaled form,
\beq
\boxed{
\frac{dp}{dt}=\frac{1}{b}\frac{de}{dt}\left\{1+\frac{(e-pb)(\gamma_a-1)}{(\gamma_a-1)e+2}\right\}
}
\label{eq:p_eqn} \, .
\eeq
Eqns.\
(\ref{eq:relmu}, 
(\ref{eq:energy_eqn} 
and
(\ref{eq:p_eqn}) represent the relativistic generalizations of the non-relativistic forms  (\ref{eq:dmudt}),
(\ref{eq:dedt}),  and
(\ref{eq:dpdt}).
In each case the relativistic correction factor, enclosed in curly brackets, reduces to unity in the non-relativistic limit $\gamma_a \rightarrow 1$.

\label{lastpage}

\end{document}